\documentclass[aps,twocolumn,superscriptaddress,showpacs,showkeys]{revtex4}
\usepackage{graphics, graphicx, dcolumn, bm,fleqn, epic, eepic}
\usepackage{amssymb, amsmath, multirow, rotate, color, float}
\usepackage{times}
\definecolor{red}{rgb}{1,0,0}
\definecolor{green}{rgb}{0,1,0}
\definecolor{blue}{rgb}{0,0,1}

\begin{document}

\title{Parameter-free resolution of the superposition of stochastic signals}

\author{Teresa Scholz}
\affiliation{Center for Theoretical and Computational Physics, University of Lisbon, Portugal}
\author{Frank Raischel}
\affiliation{Center for Geophysics, IDL, University of Lisbon, Portugal}
\affiliation{Closer Consultoria, S. Ant\~ao 7000-534, \'Evora, Portugal}
\author{Vitor V.~Lopes}
\affiliation{DEIO-CIO, University of Lisbon, Portugal}
\affiliation{Universidad de las Fuerzas Armadas-ESPE, Latacunga, Ecuador}
\author{Bernd Lehle}
\affiliation{Institute of Physics and ForWind, Carl-von-Ossietzky University of Oldenburg, Oldenburg, Germany}
\author{Matthias W\"achter}
\affiliation{Institute of Physics and ForWind, Carl-von-Ossietzky University of Oldenburg, Oldenburg, Germany}
\author{Joachim Peinke}
\affiliation{Institute of Physics and ForWind, Carl-von-Ossietzky University of Oldenburg, Oldenburg, Germany}
\author{Pedro G.~Lind}
\affiliation{Institute of Physics and ForWind, Carl-von-Ossietzky University of Oldenburg, Oldenburg, Germany}
\affiliation{Institute of Physics, University of Osnabr\"uck, Osnabr\"uck, Germany}
\date{\today}

\begin{abstract}
    This paper presents a direct method to obtain the deterministic and stochastic contribution of the sum of two independent sets of stochastic processes, one of which is composed by Ornstein-Uhlenbeck processes and the other being a general (non-linear) Langevin process.
The method is able to distinguish between all stochastic process, retrieving their corresponding stochastic evolution equations.
This framework is based on a recent approach for the analysis of multidimensional Langevin-type stochastic processes in the presence of strong measurement (or observational) noise, which is here extended to impose neither constraints nor parameters and extract all coefficients directly from the empirical data sets. Using synthetic data, it is shown that the method yields satisfactory results.
\end{abstract}

\pacs{
      02.50.Ey,  
      05.40.Ca,  
      05.45.Tp,  
      05.10.Gg,  
      }

\keywords{Stochastic Processes, Measurement Noise, Observational Noise, Signal Reconstruction, Signals Superposition}

\maketitle


\section{Introduction and motivation}

    An important topic in the analysis of time-series of complex dynamical systems is the extraction of the underlying process dynamics. Often, it is possible to reveal the deterministic and stochastic contributions of the underlying stochastic process using the It\^{o}-Langevin equation, a stochastic equation that describes the evolution of a stochastic variable. The deterministic and stochastic contributions are given by the so-called drift and diffusion coefficients, which can be directly derived from data via joint moments \cite{Friedrich1997}. This approach has been applied successfully to several areas \cite{Friedrich2011}, for example the description of turbulence \cite{Friedrich1997, Renner2001a}, the analysis of climate data \cite{Lind2005}, financial data \cite{Renner2001}, biological systems \cite{Zaburdaev2011} and wind energy production \cite{Raischel2013, Milan2013}.
    
    However, typically the time-series to be analysed is subject to noise, which is associated to the measurement devices or other sources. This so-called measurement noise, also known as observational noise, spoils the data series by hiding the underlying stochastic process. In this case, the joint moments are not accessible but only their ``noisy'' analogues. Several approaches have been published to overcome this challenge. \citet{Boettcher2006} and \citet{Lind2010} introduced a method that allows the estimation of the drift and diffusion coefficients in the presence of strong, delta-correlated Gaussian measurement noise. An alternative approach was presented by Lehle \cite{Lehle2011} that can deal with strong, exponentially correlated Gaussian noise in one dimension, which was extended to be applicable to multidimensional time-series \cite{Lehle2013}. This approach is the basis of the method presented in this paper. However, instead of using a parameterised form of the coefficients defining the stochastic process, the method extracts all coefficients directly from the data.
    
    This paper presents a methodology to distinguish between two superposed signals. In particular it presents a direct method for obtaining the evolution equation of each signal when on observes the sum of them, being one of such signals an Ornstein-Uhlenbeck (OUP) process.
If the OUP is a merely pure, uncorrelated Gaussian noise, such signal has been called ``measurement noise'' in the literarature\cite{Lehle2011,Lehle2013}, a nomenclature that we extend below to the case of correlated measurement noise, governed by an OUP.
In other words, the aim of our framework is to resolve the superposition of two stochastic processes, one of them being an OUP, which plays the role of correlated measurement noise and another being a general non-linear Langevin process. 
Specifically, we show how to extract the measurement noise parameters as well as all coefficients describing the stochastic process from the original data, to which we address below as ``noisy'' data -- as well as its associated ``noise'' statistical properties, distributions and moments -- to distinguish from the separated stochastic signals and measurement noise sources.
As we will see, the method can be applied to a set of $N$ coupled stochastic variables superposed with a set of $N$ sources of correlated measurement noise. 

The paper is structured as follows. 
The theoretical background of the Langevin analysis of stochastic processes and the extraction of the coefficients from data is briefly summarised in section~\ref{Noisystochasticprocess}. Section~\ref{Methodology} gives an overview of the method to obtain those coefficients in the presence of measurement noise. Subsequently, the two main challenges in the method are presented: a) The solution of a nonlinear equation system to obtain the measurement noise parameters, which is described in Sec.~\ref{MeasurementNoiseParameters}, and b) the solution of a system of convolution equations to estimate the joint moments of the underlying stochastic process, which is described in Sec.~\ref{Moments}. The results of application to a synthetic data set are shown in Sec.~\ref{ExampleData}, demonstrating the accuracy of the presented approach as well as its limits. Section~\ref{Conclusions} discusses possible applications of the method and concludes the paper.

\section{A general model for noisy stochastic processes} 
\label{Noisystochasticprocess}

    The evolution of a stochastic variable can be described by the It\^{o}-Langevin equation, a stochastic equation defined by a deterministic contribution (drift) and fluctuations from possible stochastic sources (diffusion). For the general case of a $N$-dimensional stochastic process $\mathbf{X}(t)$ the equation is given by: 
    \begin{equation}
            d \mathbf{x} = \mathbf{D}^{(1)}(\mathbf{x})dt +
            \sqrt{\mathbf{D}^{(2)}(\mathbf{x})}d\mathbf{W}(t), 
    \label{eq1}
    \end{equation}
    where $d\mathbf{W}$ denotes a vector of increments of independent Wiener processes with $\langle d\mathbf{W}_i\rangle = 0$ and $\langle d\mathbf{W}_i, d\mathbf{W}_j\rangle = \delta_{ij}dt$~$\forall i,j = 1,\dots,N$, where $\langle \rangle$ denotes the average and $\delta_{ij}$ the Kronecker delta. Functions $\mathbf{D}^{(1)}(\mathbf{x})$ and $\mathbf{D}^{(2)}(\mathbf{x})$ are the Kramers-Moyal coefficients of the corresponding Fokker-Planck equation that describes the evolution of the conditional probability density function and in the case the distribution of initial conditions is known one can derive the evolution equation of the joint probability density function $f(\mathbf{x}, t)$ of the stochastic variables $\mathbf{x}$. It is given by: 
    \begin{eqnarray}
        \frac{\partial f(\mathbf{x},t)}{\partial t} &=&
        -\sum_{i=1}^N\frac{\partial}{\partial x_i}
            \left [
            D_i^{(1)}(\mathbf{x})f(\mathbf{x},t)
            \right ] \cr & &+\sum_{i=1}^N\sum_{j=1}^N
            \frac{\partial ^2}{\partial x_i\partial x_j}
            \left [
            D_{i,j}^{(2)}(\mathbf{x})f(\mathbf{x},t)
            \right ] \quad .
        \label{eq2}
    \end{eqnarray}
        \begin{figure}[t]
            \includegraphics[width = 0.5\textwidth]{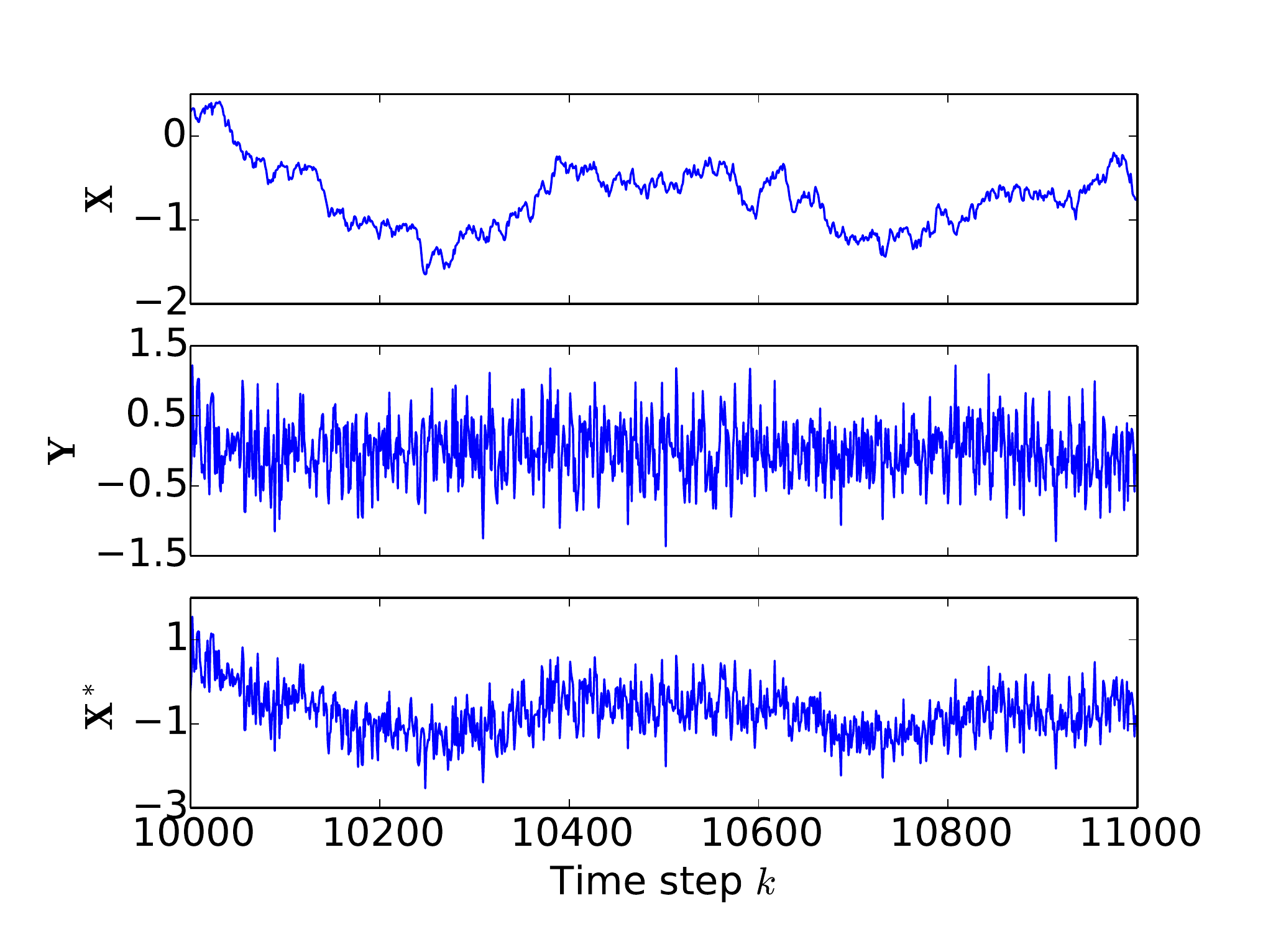}            
            \caption{A thousand values of $\mathbf{X}(t)$, the stochastic process (top), $\mathbf{Y}(t)$, the measurement noise (middle) and $\mathbf{X}^{\ast}(t)$, the noisy stochastic process (bottom).}
            \label{fig02}
        \end{figure}

    The Kramers-Moyal coefficients, also called the drift ($\mathbf{D}^{(1)}(\mathbf{x})$) and diffusion ($\mathbf{D}^{(2)}(\mathbf{x})$) coefficients, can be directly derived from measurements \cite{Friedrich2011}. 
    However, here we consider that each measured stochastic variable is the sum of two independent stochastic processes $X$ and $Y$:
    \begin{equation}
        \mathbf{X^{\ast}}(t) = \mathbf{X}(t) + \mathbf{Y}(t).
        \label{eq3} 
    \end{equation}
Since such a situation can be regarded has having a set of $N$ stochastic 
signals $\mathbf{X}(t)$ spoiled by a set of $N$ sources of measurement noise 
$\mathbf{Y}(t)$, we call the variables $\mathbf{X^{\ast}}(t)$ a $N$-dimensional
{\it noisy} stochastic process. Figure \ref{fig02} shows a specific example 
of such superposition of stochastic processes that will be addressed below
in detail.

    We assume the measurement noise $\mathbf{Y}(t)$ to be described by an Ornstein-Uhlenbeck process in $N$ dimensions: 
    \begin{equation}
        d\mathbf{y}(t) = -\mathbf{A}\mathbf{y}(t)dt + \sqrt{\mathbf{B}}d\mathbf{W}(t),
        \label{eq3prime}    
    \end{equation}
    where $\mathbf{A}$ and $\mathbf{B}$ are $N \times N$ matrices, $\mathbf{B}$ is symmetric positive semi-definite and the eigenvalues of $\mathbf{A}$ have a positive real part. Thus, the $N$-dimensional noisy stochastic process $\mathbf{X}^{\ast}$ is modeled by Eqs~(\ref{eq3}) and (\ref{eq3prime}) together. Note that here and throughout the paper $\mathbf{x}$ denotes the accessible values of any of the involved stochastic processes $\mathbf{X}(t)$,~$\mathbf{X}^{\ast}(t)$ or $\mathbf{Y}(t)$.

\section{From data to model: the inverse problem}
\label{Methodology}

\begin{figure*}[t]
        \centering
        \includegraphics[width=0.9\textwidth]{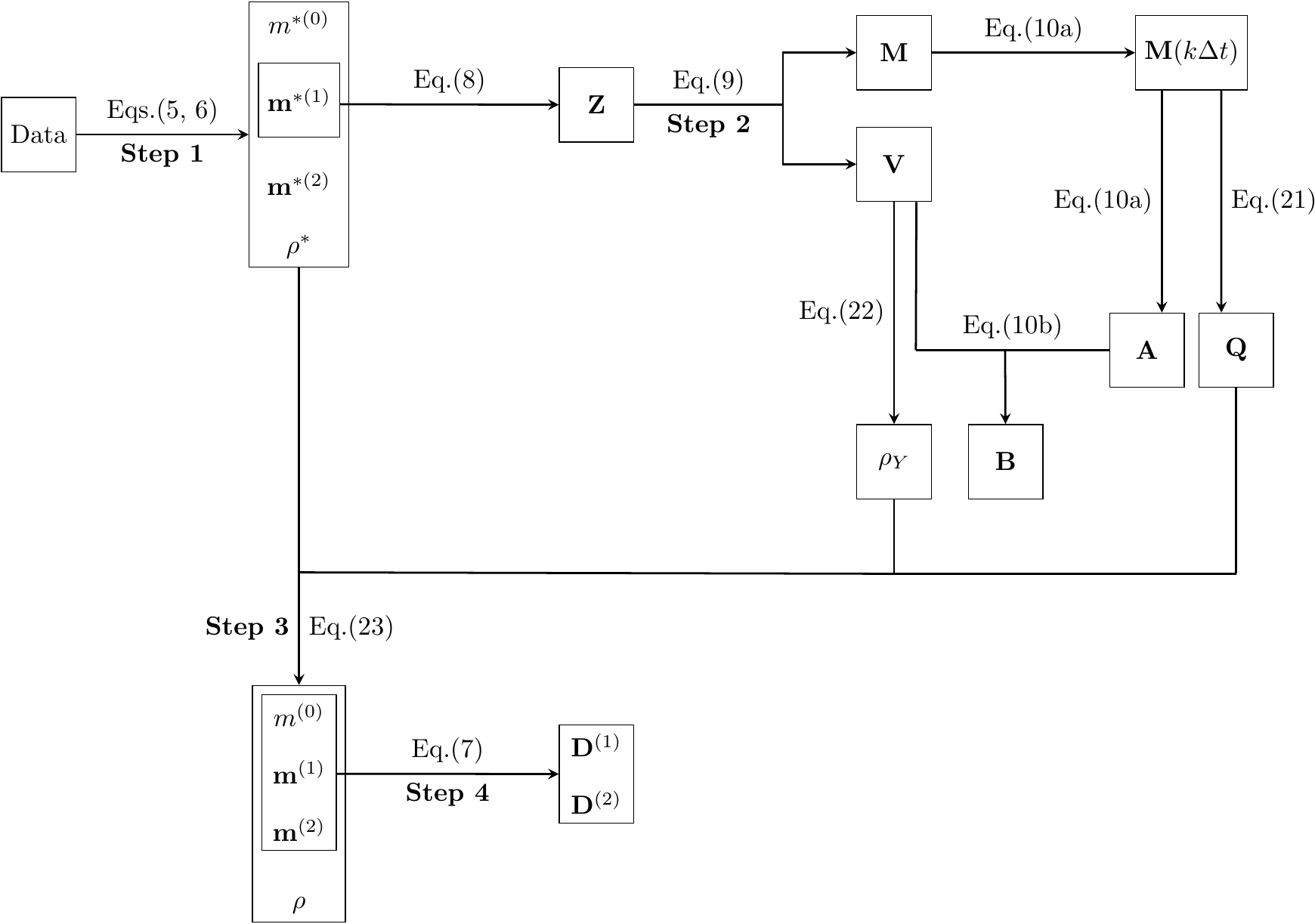}
        \caption{Graphic representation of the methodology to obtain drift and diffusion coefficents from noisy data. Functional arguments are omitted for better readability.}
        \label{fig01}
    \end{figure*}

    This section explains how to obtain the drift and diffusion coefficients along with the measurement noise parameters from noisy data $\mathbf{X}^{\ast}(t)$. The methodology is sketched in fig.~\ref{fig01} and it is based on the paper by \citet{Lehle2013} and the idea behind it is that, if the measurement noise is independent of the stochastic process, it is possible to describe the joint probability density of the noisy process as the convolution of the joint probability densities of the stochastic and the measurement noise process. This in turn allows to derive equation systems that relate the noisy moments with the noise-free moments and can be used to obtain the measurement noise parameters as well as the joint moments. Solving these equation systems in a parameter-free way is the heart of this paper and the moments can then be used to compute the drift- and diffusion coefficients. Figure \ref{fig01} shows the diagram of the four-step procedure presented in this paper.
    
    In a first step, the noisy probability density function $\rho^{\ast}(\mathbf{x}, \mathbf{x'}, \tau)$ as well as the zeroth $m^{\ast(0)}(\mathbf{x})$, first $\mathbf{m}^{\ast(1)}(\mathbf{x}, \tau)$ and second $\mathbf{m}^{\ast(2)}(\mathbf{x}, \tau)$ noisy joint moments are extracted from the data. Throughout the paper, $\tau$ refers to the time-increment which is also expressed as a multiple of the discretization time-step $\Delta t$, i.e.~$\tau = k \Delta t$ for some $k \in\mathbb{N}$. The noisy joint moments are obtained by integration through
    \begin{widetext}
        \begin{subequations}
            \begin{eqnarray}
                     m^{\ast(0)}(\mathbf{x}) &=&
                     \int_{\mathbf{x'}}\rho^{\ast}(\mathbf{x}, \mathbf{x'}, \tau)
                     dx'_1\dots dx'_N\label{eq5a}\\ 
                     m_i^{\ast(1)}(\mathbf{x}, \tau) &=& \int_{\mathbf{x'}}(x'_i(t
                     + \tau) - x_i(t))\rho^{\ast}(\mathbf{x}, \mathbf{x'}, \tau)
                     dx'_1\dots dx'_N\label{eq5b}\\ 
                     m_{i,j}^{\ast(2)}(\mathbf{x}, \tau) &=&
                     \int_{\mathbf{x'}}{ (x'_i(t+\tau) - x_i(t)) (x'_j(t + \tau) -
                       x_j(t))\rho^{\ast}(\mathbf{x}, \mathbf{x'}, \tau)}
                     dx'_1\dots dx'_N,\label{eq5c} 
            \end{eqnarray}
        \label{eq5}
        \end{subequations}
    \end{widetext}
    where $i, j = 1,\dots,N$ and  
    \begin{subequations}
        \begin{eqnarray}
              \rho^{\ast}(\mathbf{x}) &=& f(\mathbf{x}, t)\label{eq4a}\\
              \rho^{\ast}(\mathbf{x}, \mathbf{x'}, \tau) &=&
               f(\mathbf{x}, t; \mathbf{x'}, t + \tau).\label{eq4b}
        \end{eqnarray}
        \label{eq4}
    \end{subequations}

    In a second step, using the first noisy joint moments $\mathbf{m}^{\ast(1)}(\mathbf{x}, \tau)$, [Eq.~(\ref{eq5b})] the noise source is characterized by deriving its parameter matrices $\mathbf{A}$ and $\mathbf{B}$ through the solution of a nonlinear equation system. Its construction and solution are presented in section~\ref{MeasurementNoiseParameters}. 

    The numerical solutions of this equation system as well as the probability density function $\rho_Y(\mathbf{x})$ of the measurement noise are then used to formulate a system of convolution equations that relates the joint moments $m^{(0)}(\mathbf{x}), \mathbf{m}^{(1)}(\mathbf{x}, \tau)$ and $\mathbf{m}^{(2)}(\mathbf{x}, \tau)$ with their noisy analogues $\mathbf{m}^{\ast(0)}(\mathbf{x}), \mathbf{m}^{\ast(1)}(\mathbf{x}, \tau)$ and $\mathbf{m}^{\ast(2)}(\mathbf{x}, \tau)$. This third step, i.e.~the derivation and solution of this equation system is described in detail in Sec.~\ref{Moments}.
    
    In a fourth and final step, for $i,j = 1,\dots,N$, from the joint moments $m^{(0)}(\mathbf{x}), \mathbf{m}^{(1)}(\mathbf{x}, \tau)$ and $\mathbf{m}^{(2)}(\mathbf{x}, \tau)$ the drift and diffusion coefficients are computed as
    \begin{subequations}
        \begin{eqnarray}
                  D_i^{(1)}(\mathbf{x}) &=& \lim_{\tau \to 0} \frac{1}{\tau}
                  \frac{m_i^{(1)}(\mathbf{x,
                      \tau})}{m^{(0)}(\mathbf{x})}, \label{6a}\\ 
                  D_{i,j}^{(2)}(\mathbf{x}) &=& \lim_{\tau \to 0} \frac{1}{\tau}
                  \frac{m_{i,j}^{(2)}(\mathbf{x,
                      \tau})}{m^{(0)}(\mathbf{x})}.\label{eq6b}  
        \end{eqnarray}
        \label{eq6}
    \end{subequations}

\section{Extracting the measurement noise parameter matrices}
\label{MeasurementNoiseParameters}

    The measurement noise parameters $\mathbf{A}$ and $\mathbf{B}$ can be obtained by solving a nonlinear equation system. Here, a solution is approximated in the least square sense by solving an optimization problem, which is formulated making use of the symbolic framework for algorithmic differentiation and numeric optimization CasADi \citep{Andersson2010}. The optimization is performed by Ipopt, a nonlinear interior-point solver \citep{Waechter2006}. The extensive discussion of the optimization problem, i.e.~its objective function and constraints is the subject of this section. 

    \subsection{The equation system}

        From the first noisy joint moment $\mathbf{m}^{{\ast}(1)}(\mathbf{x},\tau)$ [Eq.~(\ref{eq5b})], the following $N \times N$-matrix $\mathbf{Z}$ is computed by 
        \begin{equation}
               Z_{i,j}(\tau) = \int_{\mathbf{x}}m_i^{{\ast}(1)}(\mathbf{x},\tau)x_j dx_1\dots dx_N,
        \label{eq7}
        \end{equation}
        where $i,j = 1,\dots,N$. Using $\textbf{Z}$, a non-linear equation system is constructed:
        \begin{eqnarray}
            \mathbf{Z}(k\Delta t) &=& \sum_{\nu = 1}^{\nu_{max}}\mathbf{P}^{(\nu)}(k\Delta t)^{\nu} - (\mathbf{Id} - \mathbf{M}^{k})\mathbf{V}, 
            \label{eq8}
        \end{eqnarray}
        where $\mathbf{Id}$ is the $N \times N$-identity matrix and $k \in \mathcal{K} \subset \mathbb{N}$, the set of time increments used for the estimation of the noisy moments from the data with cardinality $k_{max}$. The unknowns $\mathbf{P}^{(\nu_1)},\dots,\mathbf{P}^{(\nu_{max})}$ are auxiliary $N \times N$-matrices modeling residuals. Their definition as well as the derivation of equation system~(\ref{eq8}) are described by \citet{Lehle2013}. The unknown $N \times N$-matrices $\mathbf{V}$ and $\mathbf{M} = \mathbf{M}(\Delta t)$ are the covariance matrix and matrices of decaying correlation functions of the measurement noise, respectively, and are related to the measurement noise parameters $\mathbf{A}$ and $\mathbf{B}$ through
        \begin{subequations}
            \begin{eqnarray}
                \mathbf{M}(k \Delta t) &=& e^{-\mathbf{A}k\Delta t} = \mathbf{M}(\Delta t)^k = \mathbf{M}^k\label{eq9a}\\
                \mathbf{V} &=& \int_0^{\infty}{e^{-\mathbf{A}s}\mathbf{B}e^{-\mathbf{A}^Ts}}ds, \label{eq9b}
            \end{eqnarray}
            \label{eq9}
        \end{subequations}   
        where $\mathbf{A}^T$ denotes the transpose of matrix $\mathbf{A}$. In other words $\mathbf{A}$ is given as the matrix logarithm of $\mathbf{M}$ and $\mathbf{B}$ can be obtained through the relation $\mathbf{B} = \mathbf{VA}^T + \mathbf{AV}$ (see appendix~\ref{appendixA}). Thus, the solution of eq.~(\ref{eq8}) leads to the measurement noise parameter matrices $\mathbf{A}$ and $\mathbf{B}$. 

    \subsection{The objective function}

        To obtain a numerical estimation of the measurement noise parameters, equation system~(\ref{eq8}) is solved in the least square sense and the summed squares of the differences between its left and right hand side is minimized. This objective function is formulated making use of a lifting approach~\cite{Albersmeyer2010}, i.e.~additional variables $\mathbf{\Omega}^{(k)}$,~$k \in \mathcal{K}$, and therefore additional degrees of freedom are introduced. 
        
        The formulation of the objective function $F$ is given by
        \begin{widetext} 
            \begin{equation}
                F = \sum_{i=1}^{N}\sum_{j=1}^{N}\sum_{k \in \mathcal{K}}\left[ Z_{i,j}(k\Delta t) - \sum_{\nu = 1}^{\nu_{max}}{P_{i,j}^{(\nu)} (k\Delta t)^{\nu}} - V_{i,j} + \sum_{i' = 1}^{N}{\Omega^{(k)}_{i,i'}V_{i',j}}\right]^2
            \label{eq10}
            \end{equation}            
        \end{widetext} 
        and the $k_{max}$ constraints 
        \begin{subequations}
            \begin{align}
                \mathbf{\Omega}^{(1)} - \mathbf{M} &= 0\\
                \mathbf{\Omega}^{(k_{l + 1})} - \mathbf{\Omega}^{(k_l)}\mathbf{M}^{k_{l+1} - k_l} &= 0,
                \label{eq11}
            \end{align}
        \end{subequations}  
        where $l = 1, \dots, k_{max} - 1$, are added to the optimization problem. For sets $\mathcal{K}$ of consecutive integers, i.e.~$\mathcal{K} = \{1, \dots, k_{max}\}$, this formulation eliminates the powers of $\mathbf{M}$. For sets $\mathcal{K} = \kappa \{1, \dots, k_{max}\} = \{\kappa, \dots, \kappa k_{max}\}$,~$\kappa \in \mathbb{N}$, the powers of $\mathbf{M}$ can also be eliminated by minimizing the objective function subject to constraint~(\ref{eq11}) for $\mathbf{M}^{\prime} = \mathbf{M}^{\kappa}$ and subsequently computing the $\kappa$-th root of $\mathbf{M}^{\prime}$ using the Eigenvalue decomposition. Note that this approach might not always give sensible results, as an optimal solution $\mathbf{M}^{\prime}$ of eq.~(\ref{eq8}) subject to~(\ref{eq11}) might have negative or complex Eigenvalues. 

    \subsection{Constraints}

        The minimization of the objective function $F$ has to be performed with respect to constraints that ensure that $\mathbf{V}$ is symmetric and positive definite and $\mathbf{M}$ is stable. The constraint ensuring symmetry and positive definiteness of the matrix $\mathbf{V}$ is formulated employing the Cholesky decomposition, which is a decomposition based on a $N \times N$-lower triangular matrix with strictly positive diagonal entries. For positive definite matrices the Cholesky decomposition is unique and therefore, symmetry and positive definiteness of $\mathbf{V}$ can be imposed by 
        \begin{equation}
            \mathbf{V} = \mathbf{L}_V \mathbf{L}_V^{T},
            \label{eq12}
        \end{equation}
        where, using an exponential \textit{Ansatz}, $\mathbf{L}_V$ is given as a lower triangular matrix with positive diagonal elements by
        \begin{equation}
            \mathbf{L}_V(i,j) = 
            \begin{cases}
                0       & \quad \text{if } j > i\\
                e^{v_{i,j}}  & \quad \text{if } j = i\\
                v_{i,j} & \quad \text{if } j < i,
            \end{cases}
        \end{equation}
        with $v_{i,j} \in \mathbb{R}$.

        The matrix $\mathbf{M}(k \Delta t)$ is exponentially decaying with $k$ [Eq.~(\ref{eq9a})], which holds if and only if $\mathbf{M}$ is stable, i.e.~all its eigenvalues $\lambda_i$ are in the unit circle of the complex plane, $|{\lambda_i}| < 1$, $\forall i = 1,\dots,N$. A theorem from stability theory \cite{Martynyuk1998} states, that if there are two symmetric positive definite matrices $\mathbf{U}$ and $\mathbf{C}$ satisfying 
        \begin{equation}
            \mathbf{U} - \mathbf{M}^T \mathbf{U} \mathbf{M} = \mathbf{C}, 
        \end{equation}
        then the matrix $\mathbf{M}$ is stable. Therefore, for $\mathbf{M}$ to be stable, $\mathbf{U} - \mathbf{M}^T \mathbf{U} \mathbf{M}$ needs to be symmetric positive definite. Using the Schur complement \cite{Zhang2005} yields
        \begin{equation}
            \begin{pmatrix}
                \mathbf{U} & \mathbf{M}^T \mathbf{U}\\
                \mathbf{U M} & \mathbf{W}
            \end{pmatrix}
            \label{eq14}
        \end{equation}
        as a symmetric positive definite matrix. Thus, again using the Cholesky decomposition, the stability of $\mathbf{M}$ can be formulated as
        \begin{equation}
            \begin{pmatrix}
                \mathbf{U} & \mathbf{M}^T \mathbf{U}\\
                \mathbf{U M} & \mathbf{U}
            \end{pmatrix}
            = 
            \begin{pmatrix}
                \mathbf{L}_U & 0\\
                \mathbf{E} & \mathbf{L}_C
            \end{pmatrix}
            \begin{pmatrix}
                \mathbf{L}_U^T & \mathbf{E}^T\\
                0 & \mathbf{L}_C^T
            \end{pmatrix},
        \label{eq15}
        \end{equation}
        where $\mathbf{L}_U$, $\mathbf{L}_C$ are lower $N \times N$-triangular matrices with positive diagonal elements and $\mathbf{E}$ is a full-rank $N \times N$-matrix. Equation~(\ref{eq15}) leads to additional constraints, namely:
        \begin{subequations}
            \begin{align}
                 \mathbf{U} &= \mathbf{L}_U \mathbf{L}_U^T,      \label{eq16a}\\
                 \mathbf{M}^T\mathbf{U} &= \mathbf{L}_U \mathbf{E}^T,   \label{eq16b}\\
                 \mathbf{U} &= \mathbf{E E}^T + \mathbf{L}_C \mathbf{L}_C^T, 
            \end{align}
            \label{eq16}
        \end{subequations}
        where
            \begin{align}
                \mathbf{L}_C(i,j) &= 
                \begin{cases}
                    0       & \quad \text{if } j > i,\\
                    e^{c_{i,j}}  & \quad \text{if } j = i,\\
                    c_{i,j} & \quad \text{if } j < i,
                \end{cases} \\ 
                \mathbf{L}_U(i,j) &= 
                \begin{cases}
                    0       & \quad \text{if } j > i,\\
                    e^{u_{i,j}}  & \quad \text{if } j = i,\\
                    u_{i,j} & \quad \text{if } j < i,
                \end{cases}
            \end{align}
        with $c_{i,j}, u_{i,j} \in \mathbb{R}$.
        
        Thus, the full optimization problem is given by the minimization of $F$ [Eq.~(\ref{eq10})] subject to Eqs.~(\ref{eq11}), (\ref{eq12}) and (\ref{eq16}).

\section{Obtaining the joint moments}
\label{Moments}

    The numerical solution of equation system~(\ref{eq8}) is needed to obtain the conditional moments $m^{(0)}(\mathbf{x}), \mathbf{m}^{(1)}(\mathbf{x}, \tau)$ and $\mathbf{m}^{(2)}(\mathbf{x}, \tau)$. Using $\mathbf{M}$, the auxiliary $N \times N$-matrix $\mathbf{Q}(\tau)$ is computed by 
    \begin{equation}
        \mathbf{Q}(\tau) = \mathbf{Q}(k\Delta t) = (\mathbf{Id} - \mathbf{M}(k \Delta t))\mathbf{V}
        \label{eq18}
    \end{equation}
    and Eq.~(\ref{eq9a}). Assuming, that the measurement noise is distributed with a normalized Gauss function in $\mathbf{x}$ with covariance $\mathbf{V}$, it is possible to compute the probability density function $\rho_Y(\mathbf{x})$ of $\mathbf{Y}(t)$ by
    \begin{equation}
        \rho_Y(\mathbf{x}) = \frac{1}{\sqrt{(2\pi)^N|\det(\mathbf{V})|}} e^{\frac{1}{2}\mathbf{x}^T\mathbf{V}^{-1}\mathbf{x}}  \, .
        \label{eq19}
    \end{equation}

    \begin{widetext}
    Using $\mathbf{Q}(\tau)$ given by Eq.~(\ref{eq18}) and $\rho_Y(\mathbf{x})$ given by Eq.~(\ref{eq19}), it is possible to relate the conditional moments $m^{(0)}(\mathbf{x}), \mathbf{m}^{(1)}(\mathbf{x}, \tau)$ and $\mathbf{m}^{(2)}(\mathbf{x}, \tau)$ with their noisy analogues $m^{\ast(0)}(\mathbf{x}), \mathbf{m}^{\ast(1)}(\mathbf{x}, \tau)$ and $\mathbf{m}^{\ast(2)}(\mathbf{x}, \tau)$:
        \begin{subequations}
            \begin{align}
                m^{\ast(0)}(\mathbf{x}) &= \rho_Y(\mathbf{x}) \ast m^{(0)}(\mathbf{x}),\label{eq20a}\\
                m_i^{\ast(1)}(\mathbf{x}, \tau) &= \rho_Y(\mathbf{x}) \ast m_i^{(1)}(\mathbf{x}, \tau) + \sum_{i'}Q(\tau)_{i,i'}\partial_{i'}m^{\ast(0)}(\mathbf{x}),\label{eq20b}\\
                m_{i,j}^{\ast(2)}(\mathbf{x}, \tau) &= \rho_Y(\mathbf{x}) \ast m_{i,j}^{(2)}(\mathbf{x}, \tau) + (Q(\tau)_{i,j} + Q(\tau)_{j,i})m^{\ast(0)}(\mathbf{x})  
                + \sum_{i'}Q(\tau)_{i,i'}\partial_{i'}m_j^{\ast(1)}(\mathbf{x}, \tau) \nonumber\\ 
                &+ \sum_{j'} Q(\tau)_{j,j'}\partial_{j'}m_i^{\ast(1)}(\mathbf{x}, \tau) - \sum_{i'}\sum_{j'}Q(\tau)_{i,i'}Q(\tau)_{j,j'}\partial_{i'}\partial_{j'} m^{\ast(0)}(\mathbf{x}), \label{eq20c}        
            \end{align}
        \label{eq20}
        \end{subequations}
    where $\ast$ represents the convolution operator and $i,j = 1, \dots,N$. Solving Eqs.~(\ref{eq20}) yields the joint moments $m^{(0)}(\mathbf{x})$, $\mathbf{m}^{(1)}(\mathbf{x}, \tau)$ and $\mathbf{m}^{(2)}(\mathbf{x}, \tau)$. The formulation of the corresponding optimization problems is presented in this section. 

    \subsection{The objective functions}
        
        Equations~(\ref{eq20}) are solved one by one and for each component separately in the least square sense. Therefore, for $i,j = 1, \dots\,N$, the objective functions are given by
            \begin{subequations}
                \begin{align}
                    F^{(0)} &= \sum_{\mathbf{x}} \vert m^{\ast (0)}(\mathbf{x}) - \rho_Y(\mathbf{x}) \ast m^{(0)}(\mathbf{x})\vert^2 , \\
                    F_i^{(1)} &= \sum_{\mathbf{x}} \vert m_i^{\ast (1)}(\mathbf{x}, \tau) - \rho_Y(\mathbf{x}) \ast m^{(1)}(\mathbf{x}, \tau) - \sum_{i'}Q(\tau)_{i,i'}\partial_{i'}m^{\ast(0)}(\mathbf{x})\vert^2 , \\
                    F_{i,j}^{(2)} &= \sum_{\mathbf{x}} \vert m_{i,j}^{\ast(2)}(\mathbf{x}, \tau) - \rho_Y(\mathbf{x}) \ast m_{i,j}^{(2)}(\mathbf{x}, \tau) - (Q(\tau)_{ij} + Q(\tau)_{j,i}) m^{\ast(0)}(\mathbf{x}) - \sum_{i'}Q(\tau)_{i,i'}\partial_{i'}m_j^{\ast(1)}(\mathbf{x}, \tau) \nonumber\\
                    &- \sum_{j'} Q(\tau)_{j,j'}\partial_{j'}m_i^{\ast(1)}(\mathbf{x}, \tau) + \sum_{i'}\sum_{j'}Q(\tau)_{i,i'}Q(\tau)_{j,j'}\partial_{i'}\partial_{j'} m^{\ast(0)}(\mathbf{x})\vert^2. 
                \label{eq21} 
                \end{align}            
            \end{subequations}
        \end{widetext}
           
    \subsection{Regularization}

        Equations~(\ref{eq20}) are multidimensional integral equations of the convolution type, which are known to be ill-posed \cite{Zhang2015} and to deal with this, the mathematical technique of regularization has been employed. 

The idea behind it is to introduce additional information in order to solve the problem, here, a restriction for smoothness of the moments was chosen to compensate numerical fluctuations that occur due to binning.

        Therefore, a weighted penalty term given by
        \begin{subequations}
            \begin{align}
                p^{(0)} &= \sum_{\mathbf{x}} \sum_{l = 1}^N (\frac{\partial m^{(0)}(\mathbf{x})}{\partial x_{l} })^2 , \\
                p_i^{(1)}(\tau) &= \sum_{\mathbf{x}} \sum_{l = 1}^N (\frac{\partial m_i^{(1)}(\mathbf{x}, \tau)}{\partial x_{l} })^2 , \\
                p_{i,j}^{(2)} (\tau) &= \sum_{\mathbf{x}} \sum_{l = 1}^N (\frac{\partial m_{i,j}^{(2)}(\mathbf{x}, \tau)}{\partial x_{l} })^2,
            \end{align}
        \label{eq22}
        \end{subequations}
        is added to the corresponding objective function.

        The weight $\alpha^{(0)}, \alpha_i^{(1)}, \alpha_{i,j}^{(2)}$ with $i,j = 1,\dots,N$, strongly influences the outcome of the optimization problem and therefore it is crucial to select an appropriate weight. The strategy employed in this paper is described in the following.

        The zeroth joint moment $m^{(0)}(\mathbf{x})$ is known to be constant in $\tau$ and all components of the first and second joint moments, $m_i^{(1)}(\mathbf{x}, \tau)$ and $m_{i,j}^{(2)}(\mathbf{x}, \tau)$ respectively, are known to be linear in $\tau$\cite{Lehle2011,Friedrich2011}. Approximations of each of the joint moments $m^{\prime(0)}(\mathbf{x}, \alpha^{(0)})$, $m_i^{\prime(1)}(x, \tau, \alpha_i^{(1)})$ and $m_{i,j}^{\prime(2)}(x, \tau, \alpha_{i,j}^{(2)})$ are computed for several time increments $\tau = k \Delta t$ with $k = 1, \dots, k_{max}$ and several weights $\alpha^{(0)}, \alpha_i^{(1)}, \alpha_{i,j}^{(2)} \in \mathcal{A} = \{\alpha_{min},\dots,\alpha_{max}\}$. For each penalty weight a linear fit is performed on the numerical solutions in $\tau$, for each $\mathbf{x}$ yielding a straight line $g(\mathbf{x}, \alpha, \tau) = a(\mathbf{x}, \alpha) + b(\mathbf{x}, \alpha) \tau$, for $\alpha \in \{\alpha^{(0)}, \alpha_i^{(1)}, \alpha_{ij}^{(2)}\}$.

        Since the zeroth joint moment $m^{(0)}(\mathbf{x})$ is constant, the corresponding slopes $b(\mathbf{x}, \alpha^{(0)})$ should vanish for the appropriate penalty weight. Therefore, to choose the weight for the approximation of the zeroth moment the sum of the absolute values of the slopes over all $\mathbf{x}$ is computed as
        \begin{equation}
            s^{(0)}(\alpha^{(0)}) = \sum_{\mathbf{x}} \vert b(\mathbf{x}, \alpha^{(0)}) \vert \, .
        \end{equation}

        For the first and second moment that are linear in $\tau$, the squared residuals are computed, weighted with the absolute value of the corresponding slope and summed up over all $\mathbf{x}$:
        \begin{widetext}
        \begin{subequations}
            \begin{align}
                r_i^{(1)}(\mathbf{x}, \alpha_i^{(1)}, \tau) &= \sum_{\mathbf{x}} \vert b(\mathbf{x}, \alpha_i^{(1)}) \vert (g(\mathbf{x}, \alpha_i^{(1)}, \tau) - m_i^{\prime(1)}(\mathbf{x}, \alpha_i^{(1)}, \tau))^2,\\
                r_{i,j}^{(2)}(\mathbf{x}, \alpha_{i,j}^{(2)}, \tau) &= \sum_{\mathbf{x}} \vert b(\mathbf{x}, \alpha_{i,j}^{(2)}) \vert (g(\mathbf{x}, \alpha_{i,j}^{(2)}, \tau) - m_{i,j}^{\prime(2)}(\mathbf{x}, \alpha_{i,j}^{(2)}, \tau))^2,
            \end{align}
        \end{subequations}
        \end{widetext}
        where $i,j = 1,\dots,N$.
        
        Choosing the appropriate weight for the penalty term is a trade-off between minimising the summed slopes (zeroth joint moment) and the weighted residuals (first and second joint moment) of the linear fit and minimising the difference of left and right hand side of Eqs.~(\ref{eq8}), which increases after a threshold weight for each of the joint moments: Since an increasing penalty weight leads to flatter moments, both the summed slopes and the summed residuals decrease with increasing weight. If the penalty weight is too large, the penalty term outweighs the corresponding objective function term $F^{(0)}, F_i^{(1)}$ and $F_{i,j}^{(2)}$. As a consequence, the resulting approximated joint moment does not resemble the empirical joint moment. Therefore, to pick an appropriate weight for the estimation of $m^{(0)}(\mathbf{x})$, the sum of $F^{(0)}$ and $s^{(0)}(\alpha^{(0)})$ is minimized. Typically, for the first and second joint moments, the sum of the weighted residuals $r_i^{(1)}(\mathbf{x}, \alpha_i^{(1)}, \tau)$,~$r_{i,j}^{(2)}(\mathbf{x}, \alpha_{i,j}^{(2)}, \tau)$, and the objective function terms $F_i^{(1)}$, $F_{i,j}^{(2)}$ are not of the same order of magnitude. Therefore, they are normalized by their maximum value and the offset in $F_i^{(1)}$, $F_{i,j}^{(2)}$ is removed. The appropriate weight is given by the one corresponding to the minimum of the sum of the objective function with the residual.

    \subsection{The final optimization problems}

        Since $m^{(0)}(\mathbf{x})$ is a probability density function (defined in analogy to its noisy counterpart [see Eq.~(\ref{eq5a})], its integral over the full range of $\mathbf{x}$-values equals one. Therefore, the constraint
        \begin{equation}
            \int_{\mathbf{x}} m^{(0)}(\mathbf{x}) dx_1 \dots dx_N = 1
            \label{eq29}
        \end{equation}
        is added to the optimization problem for the estimation of $m^{(0)}(\mathbf{x})$. The full optimization problem is therefore given by the minimization of $F^{(0)} + \alpha^{(0)} p^{(0)}$, subject to Eq.~(\ref{eq29}) with positive real $\alpha^{(0)}$.

        For the first and second moments no additional constraints are imposed and the final optimization problems are simply given by the minimization of $F_i^{(1)} + \alpha_i^{(1)} p_i^{(1)}$ and $F_{i,j}^{(2)} + \alpha_{i,j}^{(2)} p_{i,j}^{(2)}$ with positive real $\alpha_i^{(1)}, \alpha_{i,j}^{(2)}$ and $i,j = 1,\dots,N$.

\section{An illustrative example}
\label{ExampleData}

    In the previous two sections this papers main focus of obtaining the measurement noise parameters $\mathbf{A}$ and $\mathbf{B}$ as well as the joint moments $m^0(\mathbf{x})$, $\mathbf{m}^1(\mathbf{x}, \tau)$ and $\mathbf{m}^2(\mathbf{x}, \tau)$, was described. To demonstrate that this framework yields correct results it is tested on two-dimensional synthetic data, which is described in this section. Moreover, the results of solving equation systems~(\ref{eq8}) and~(\ref{eq20}) for this example are presented and discussed.

    \subsection{The data}

        Two time-series, each comprising $10^6$ data points are generated: $\mathbf{X}(t)$, the stochastic process and $\mathbf{Y}(t)$, the measurement noise. Together they yield the noisy stochastic process $\mathbf{X}^{\ast}(t)$ [see Eq.~(\ref{eq3})]. The time-step between consecutive datapoints is $\Delta t = 0.005$ in arbitrary units.

        The time-series of the stochastic process $\mathbf{X}(t)$ is obtained by stochastic integration of Eq.~(\ref{eq1}), where, with $\mathbf{x} = (x_1, x_2)$
        \begin{equation*}
            \mathbf{D}^{(1)}(\mathbf{x}) =  
            \begin{pmatrix}
            x_1 - x_1x_2\\
            x_1^2 - x_2
            \end{pmatrix},
            \qquad
            \mathbf{D}^{(2)}(\mathbf{x}) =
            \begin{pmatrix}
            0.5 & 0\\
            0 & 0.5(1 + x_1^2)
            \end{pmatrix}.
        \end{equation*}

        The time-series of the measurement noise $\mathbf{Y}$ is obtained by stochastic integration of Eq.~(\ref{eq3prime}), where 
        \begin{equation*}
            \mathbf{A} =  
            \begin{pmatrix}
            200 & -\frac{200}{3}\\
            \\
            0 & \frac{200}{3}
            \end{pmatrix},
            \qquad
            \mathbf{B} =
            \begin{pmatrix}
            75 & -\frac{425}{12}\\
            \\
            -\frac{425}{12} & \frac{125}{6}
            \end{pmatrix}.
        \end{equation*}

        From $\mathbf{A}$ and $\mathbf{B}$, the values of the correlation matrix $\mathbf{M}$ and the covariance matrix $\mathbf{V}$ are computed as 
        \begin{equation*}
            \mathbf{M} =  
            \begin{pmatrix}
                e & \frac{1}{2}(e^{-\frac{1}{3}} - e^{-1})\\
                \\
                0 & e^{-\frac{1}{3}}
            \end{pmatrix},
            \qquad
            \mathbf{V} =
            \begin{pmatrix}
                \frac{5}{32} & -\frac{3}{32}\\
                \\
                -\frac{3}{32} & \frac{5}{32}
            \end{pmatrix},
        \end{equation*}
        see Eq.~(\ref{eq9}). To illustrate the time-series, fig.~\ref{fig02} shows $\mathbf{X}(t)$ (top), $\mathbf{Y}(t)$ (middle) and $\mathbf{X}^{\ast}(t)$ (bottom) for $1000$ time-steps.

        The ratio $\frac{\sigma_{noise}}{\sigma_{process}}$ between the standard deviation of the measurement noise $\sigma_{noise}$ and the standard deviation of the stochastic process $\sigma_{process}$ is approximately $0.4$ for the first and $0.5$ for the second component. For the estimation of the noisy moments $m^{\ast (0)}(\mathbf{x})$, $\mathbf{m}^{\ast (1)}(\mathbf{x}, \tau)$ and $\mathbf{m}^{\ast (2)}(\mathbf{x}, \tau)$ time-steps $\tau = k \Delta t$ with $k = 1, 2, 3, 4, 5, 10, 20, 30, 40, 50$ were chosen.

    \subsection{Measurement noise parameters}

        Equation system~(\ref{eq8}) depends on the number of time-steps $k$ included into the system and the degree $\nu_{max}$ of the polynomial in $\mathbf{P}$. A determined or overdetermined system is obtained when $\nu_{max} \leq k_{max} - 2$. Therefore, to choose an appropriate $\nu_{max}$ for a set $\mathcal{K}$ of time-steps $k$, the system is solved for different values of $\nu_{max}$ up to $k_{max} - 2$ and the values corresponding to the minimal error are chosen. To quantify the error, the $2$-norm, $D_M$ and $D_V$ respectively, between the eigenvalues of $\mathbf{M}$ and $\mathbf{V}$ and their approximations is used. To apply the method described in Sec.~\ref{MeasurementNoiseParameters}, two sets of time-steps involving different time-scales were used, $\mathcal{K}_1 = \{1, 2, 3, 4, 5\}$ and $\mathcal{K}_2 = 5\mathcal{K}_1$. The results are presented in table~\ref{table1}.
        \begin{figure}[t]
            \centering
            \includegraphics[width = 0.5\textwidth]{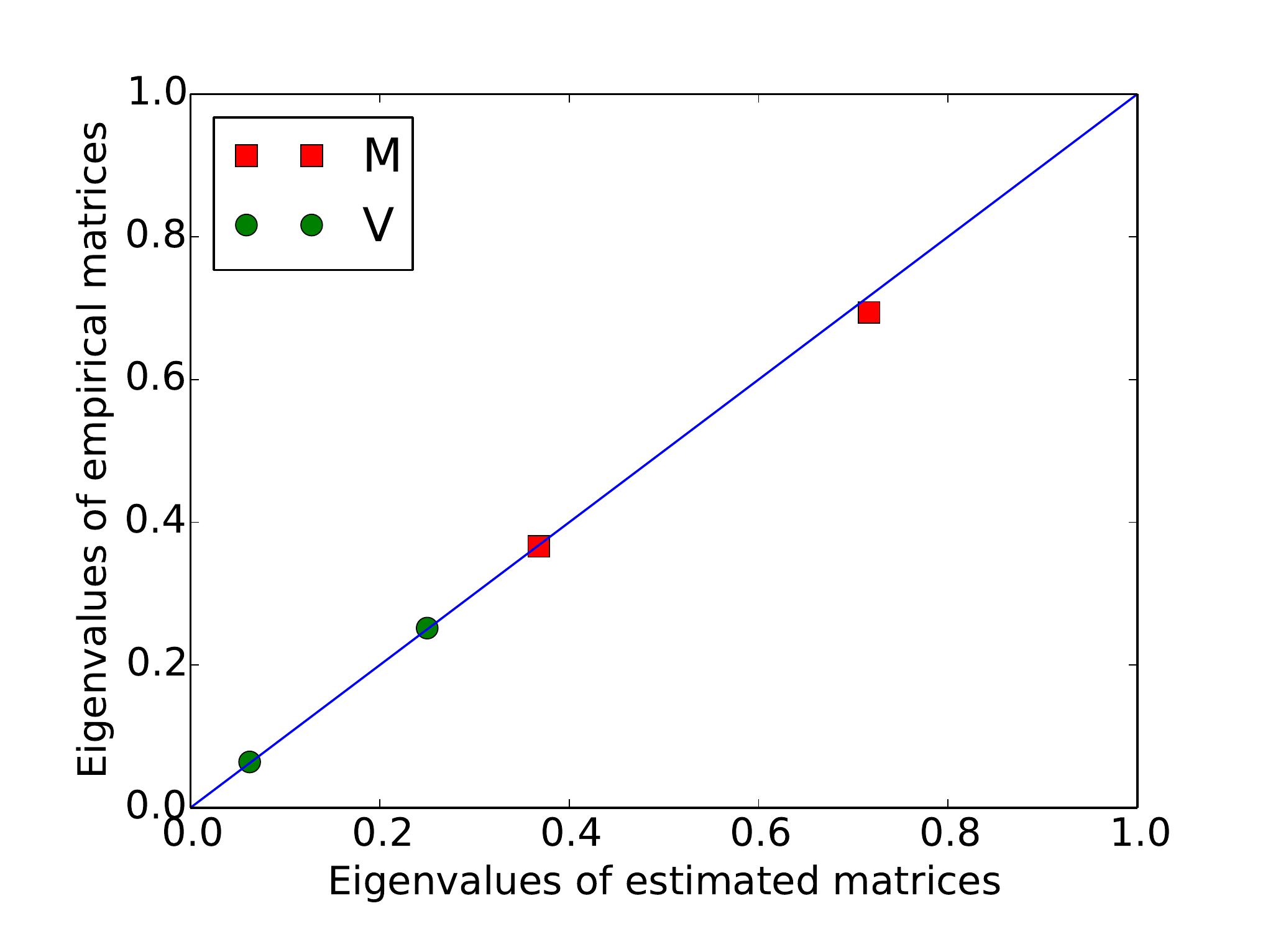}
            \caption{Eigenvalues of the measurement noise covariance matrix $\textbf{V}$ and its matrix of correlation functions $\textbf{M}$, computed analytically and by solving eq.~(\ref{eq8}) in the least square sense.}
        \label{fig03}
        \end{figure}        
        \begin{table}[tb]
            \begin{tabular}{| c c | c  c  c |}
                \hline
                & $\nu_{max}$ & 1 & 2 & 3 \\ \hline
                \multirow{2}{*}{$\mathcal{K}_1$} 
                 & $D_M$ & 0.024 & 0.016 & 0.071\\ 
                 & $D_V$  & 0.262 & 0.265 & 0.908\\
                \hline   
                \multirow{2}{*}{$\mathcal{K}_2$} 
                 & $D_M$ & 0.026 & 0.011 & 0.168\\ 
                 & $D_V$  & 0.001 & 0.004 & 0.017\\
                \hline
            \end{tabular}
            \caption{Distance $D_M$ and $D_V$, respectively, between the eigenvalues of $\mathbf{M}$ and $\mathbf{V}$ and their approximations for two sets of time-steps $\mathcal{K}_1$ and~$\mathcal{K}_2$ and different values of $\nu_{max}$.}
            \label{table1}
        \end{table}

        Table \ref{table1} shows, that the best approximation is obtained for $\mathcal{K}_2$ and $\nu_{max} = 2$. For both sets $\mathcal{K}_1$ and~$\mathcal{K}_2$ the worst results are obtained for $\nu_{max} = 3$. However, for all other cases the optimization yields good results, showing that the method is robust with regard to the time scale of $\mathcal{K}$.

        Figure \ref{fig03} shows the eigenvalues of the matrices $\mathbf{M}$ and $\mathbf{V}$ computed by the optimization (with $\mathcal{K}_2$ and $\nu_{max} = 2$) plotted against their analytical analogues on the left. For comparison, the bisectrix is plotted in blue, indicating perfect agreement between simulation and reconstructed values. 
        \begin{figure}[hbt]
            \includegraphics[width = 0.5\textwidth]{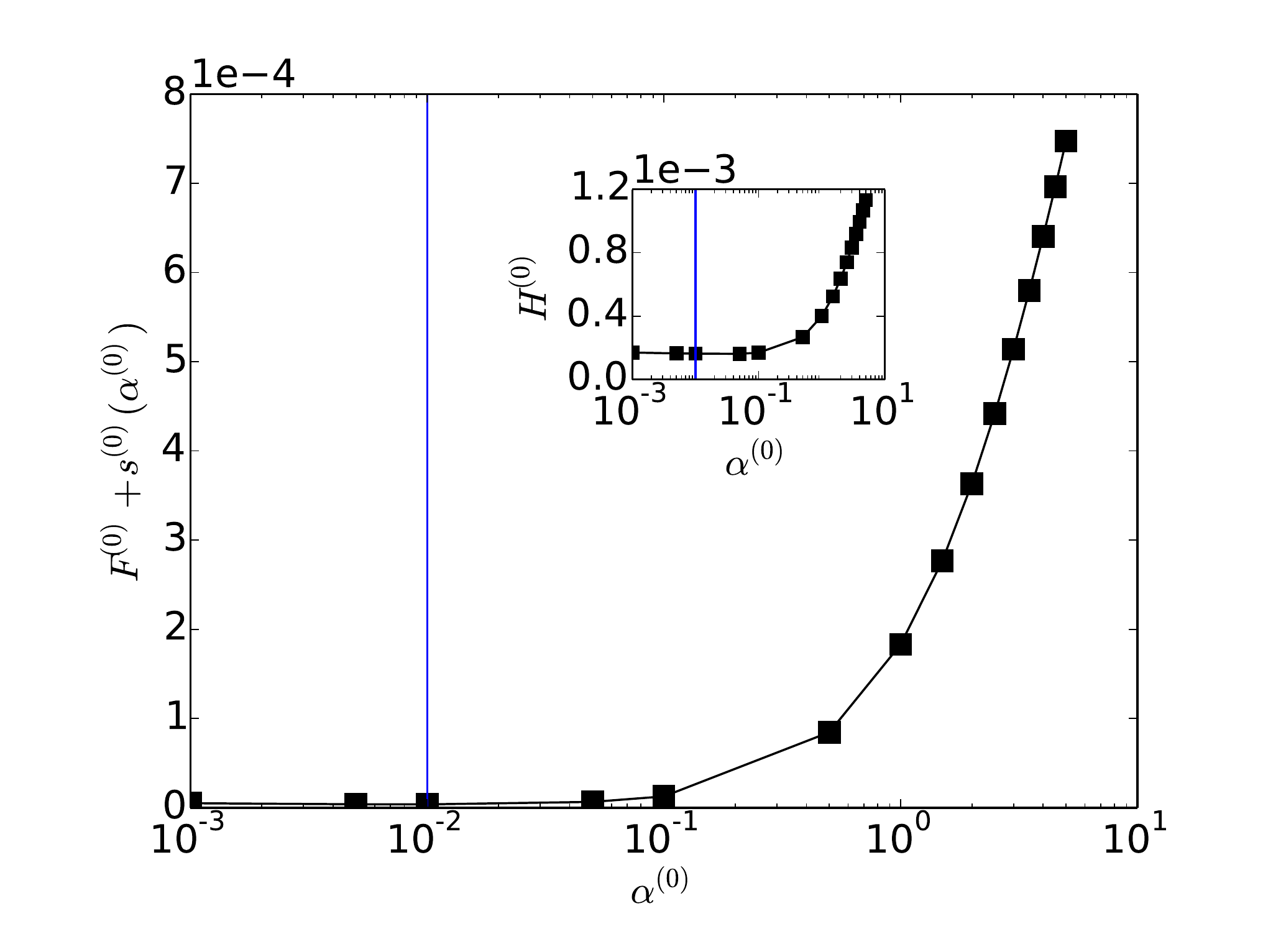}
            \includegraphics[width = 0.5\textwidth]{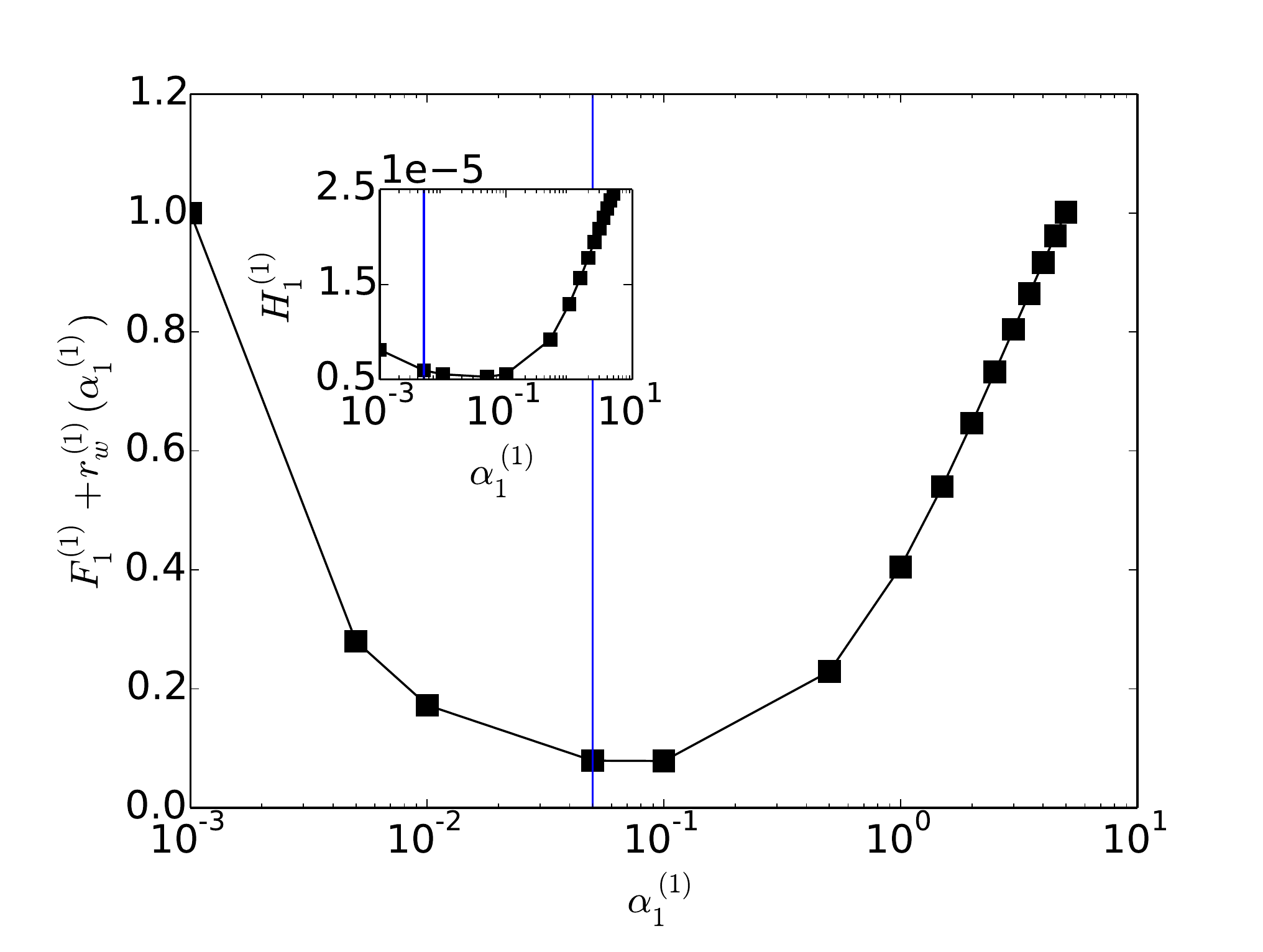}
            \caption{Illustration of the figures indicating the penalty weight: sum of $F^{(0)}$ and $s^{(0)}(\alpha^{(0)})$ (top) and $r_i^{(1)}(\mathbf{x}, \tau, \alpha_i^{(1)})$ and $F_i^{(1)}$ (bottom) for the approximations $m^{\prime(0)}(x, \alpha_0)$ (top) and $m_i^{\prime(1)}(x, \tau, \alpha_1)$ (bottom) in dependence of the penalty weight. The inset shows the difference of the estimated and empirical zeroth (top) and first (bottom) joint moment. For the second moments $m_{ij}^{\prime(2)}$ results similar to $m_i^{\prime(1)}$ are obtained.}
            \label{fig04}
        \end{figure}
        \begin{figure*}[htb]
            \includegraphics[width = 0.49\textwidth]{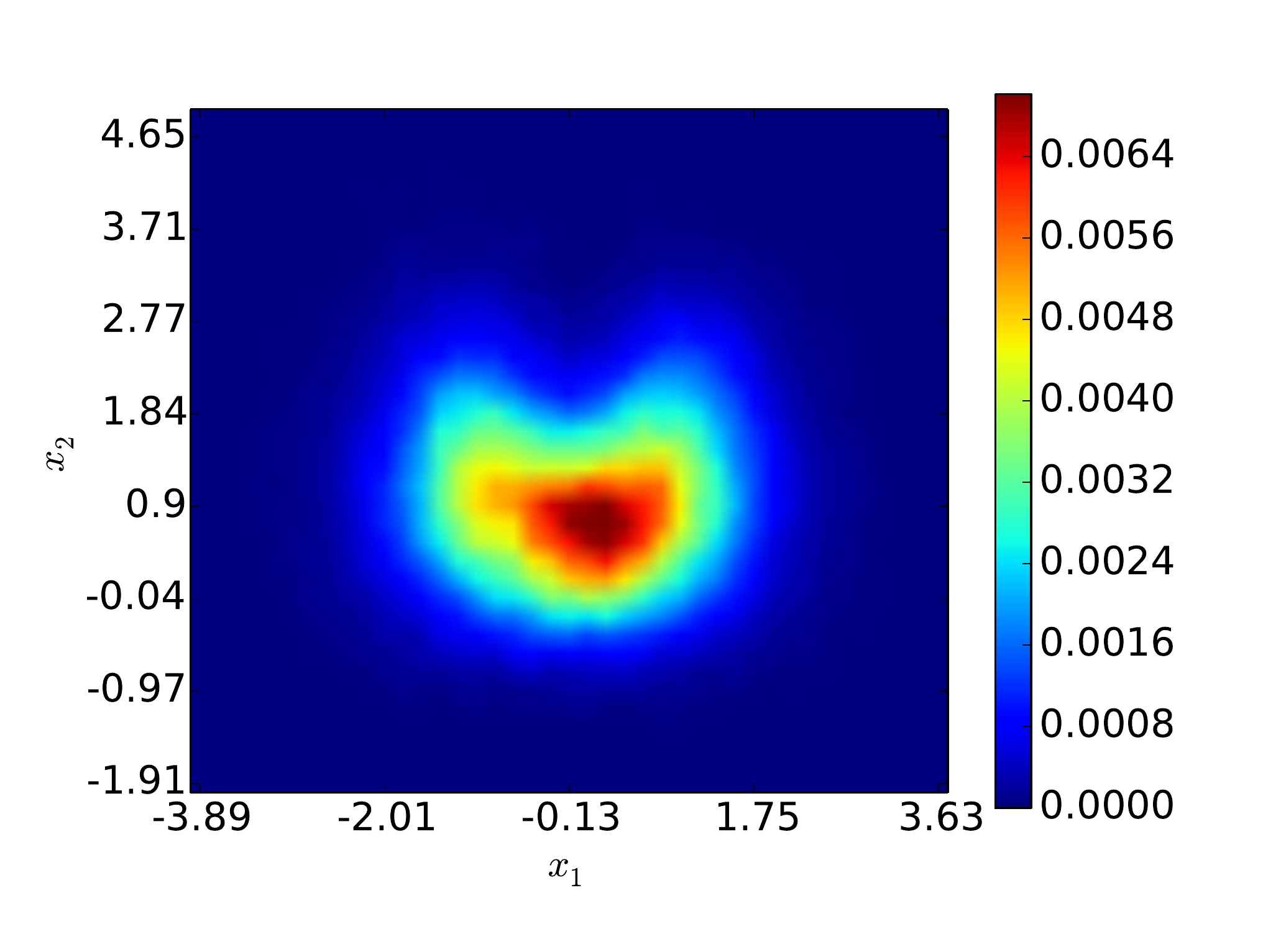}%
            \includegraphics[width = 0.49\textwidth]{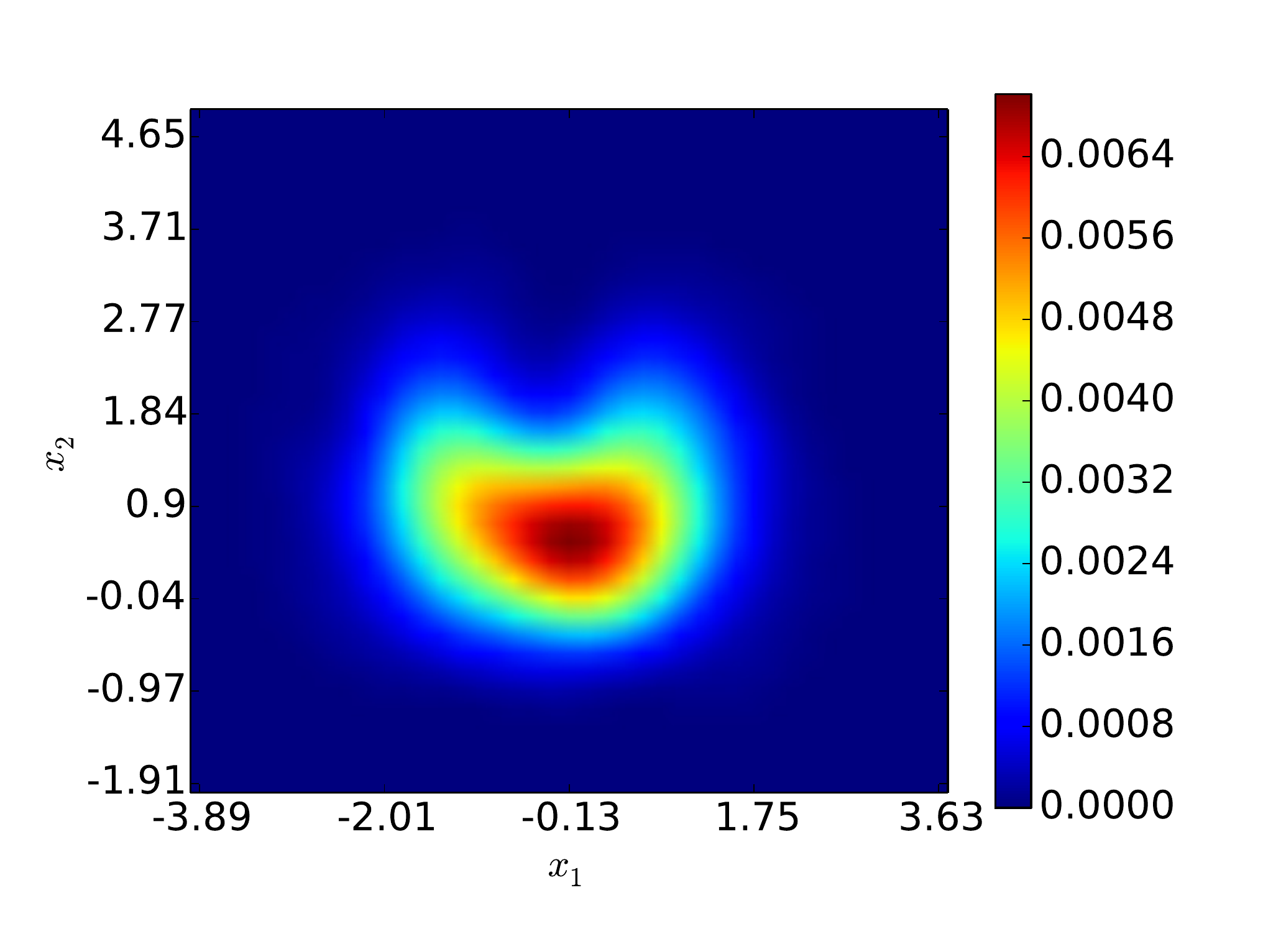}          
            \includegraphics[width = 0.49\textwidth]{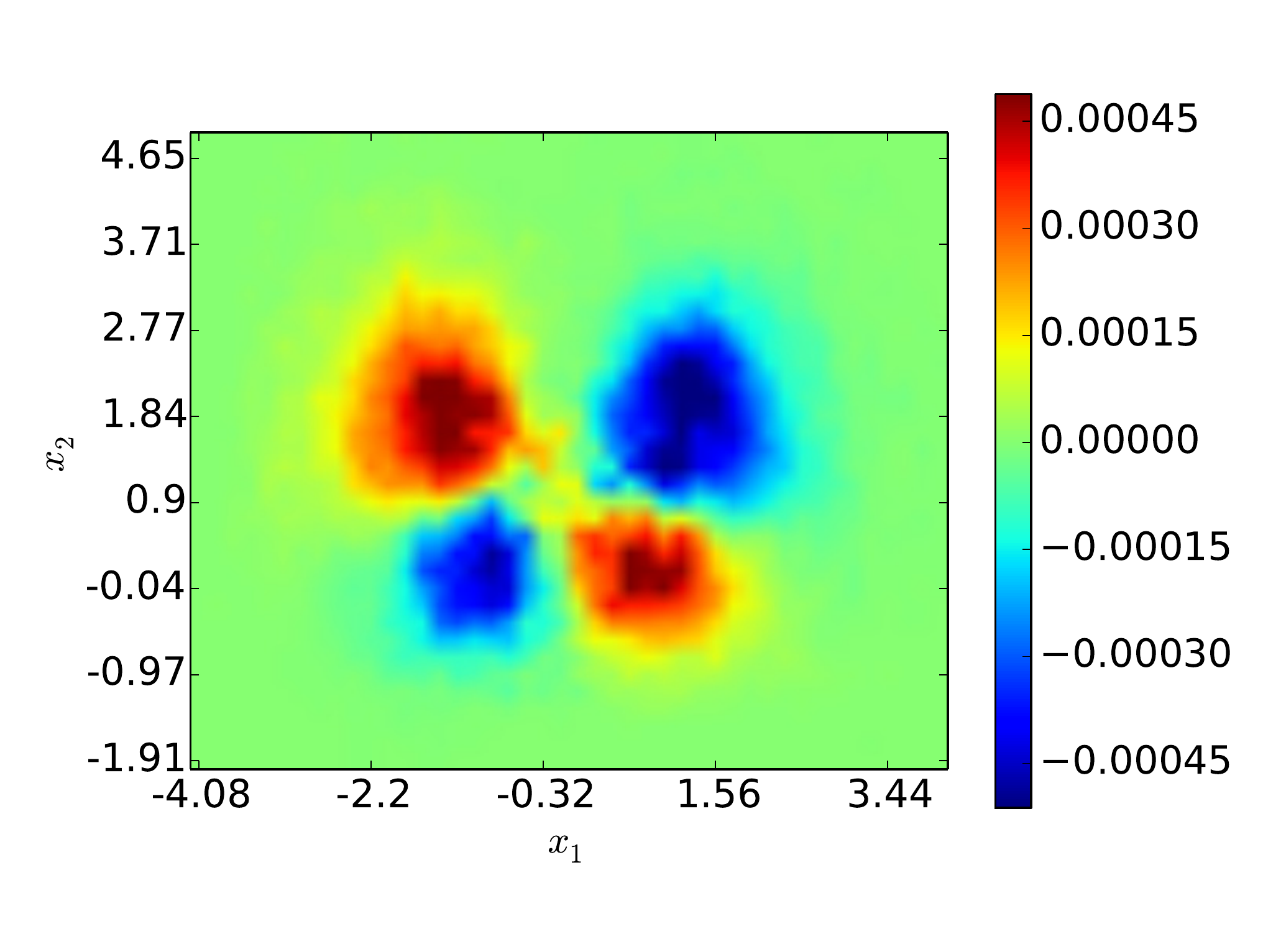}%
            \includegraphics[width = 0.49\textwidth]{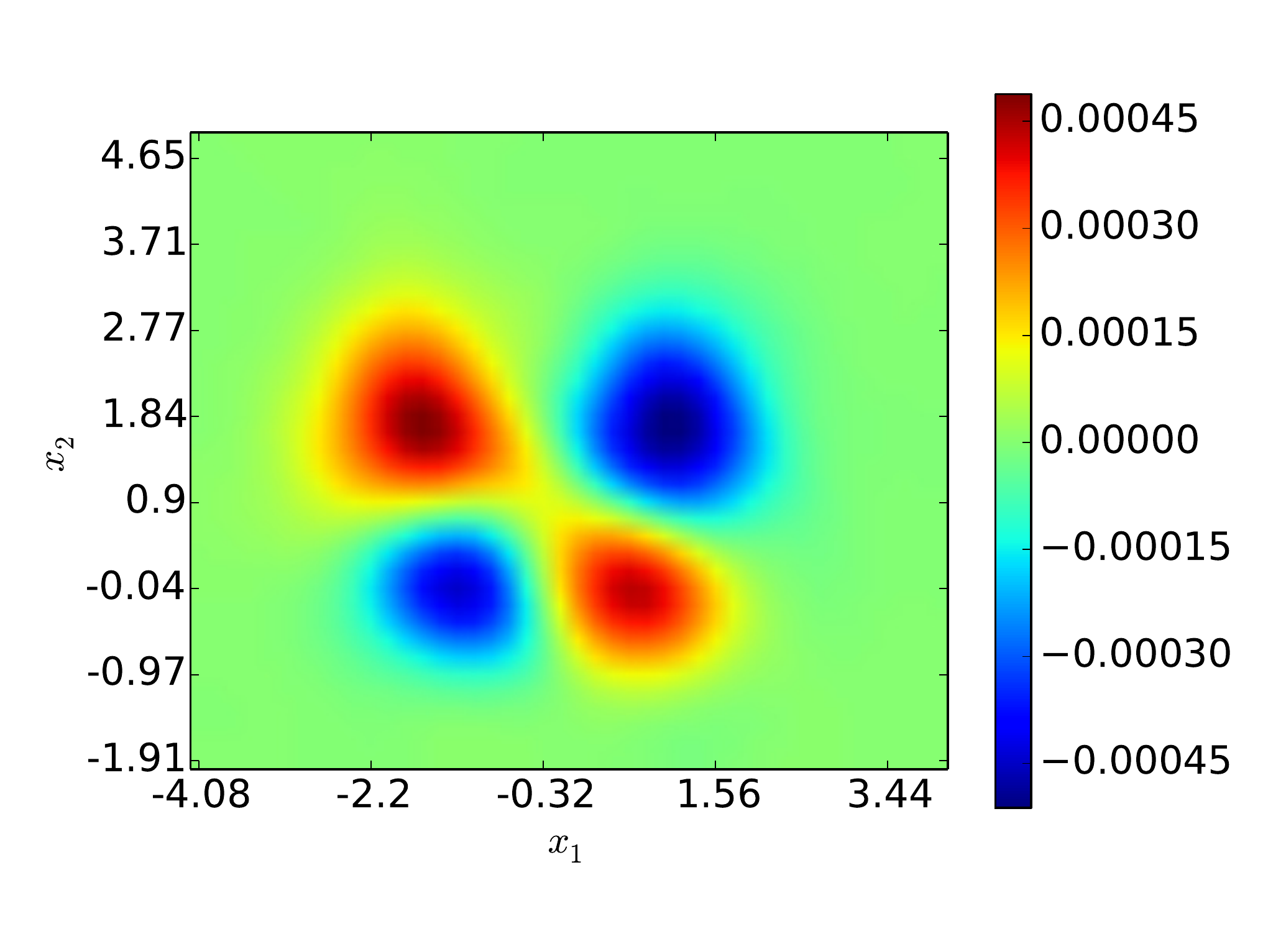}
            \includegraphics[width = 0.49\textwidth]{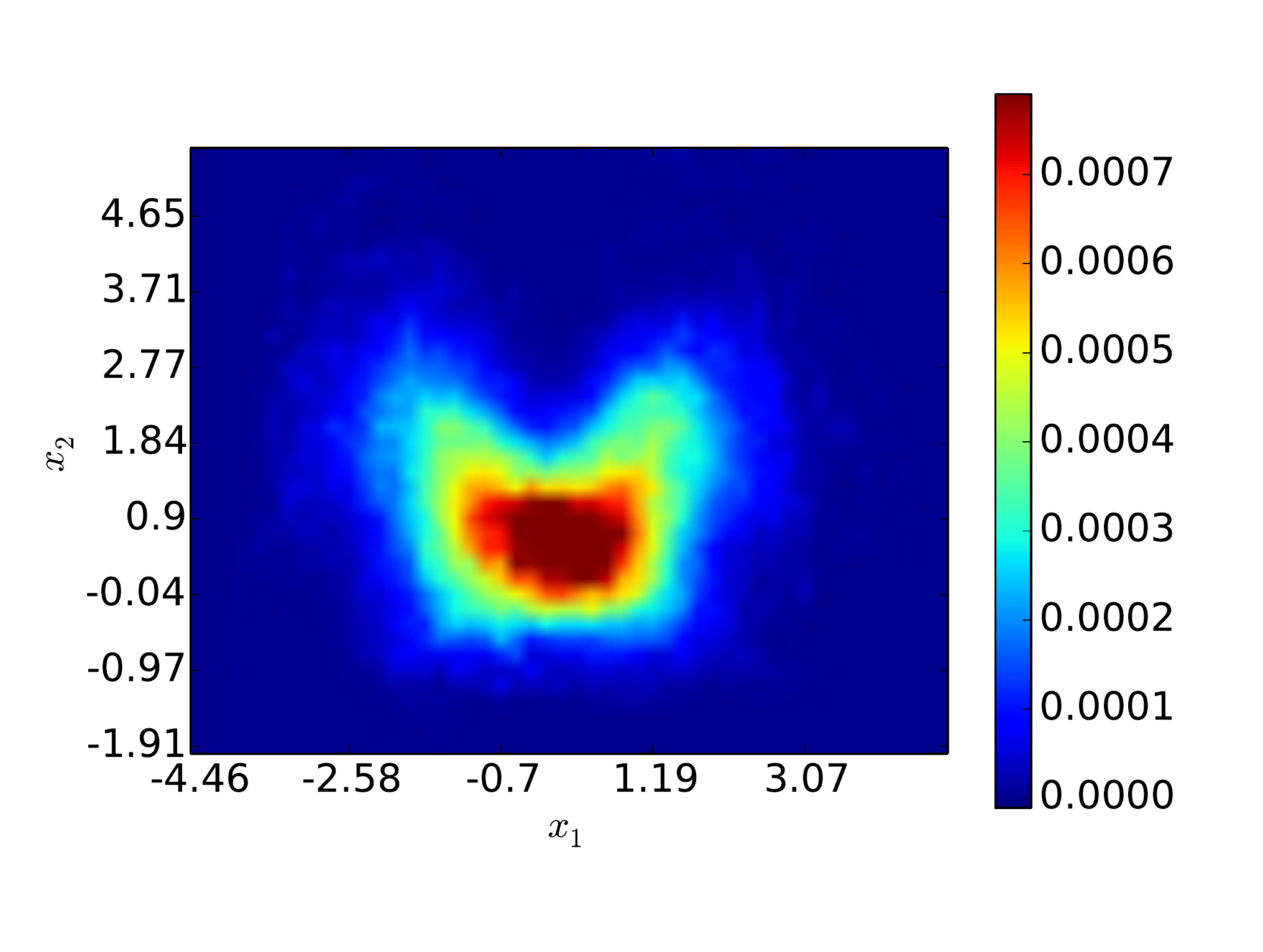}%
            \includegraphics[width = 0.49\textwidth]{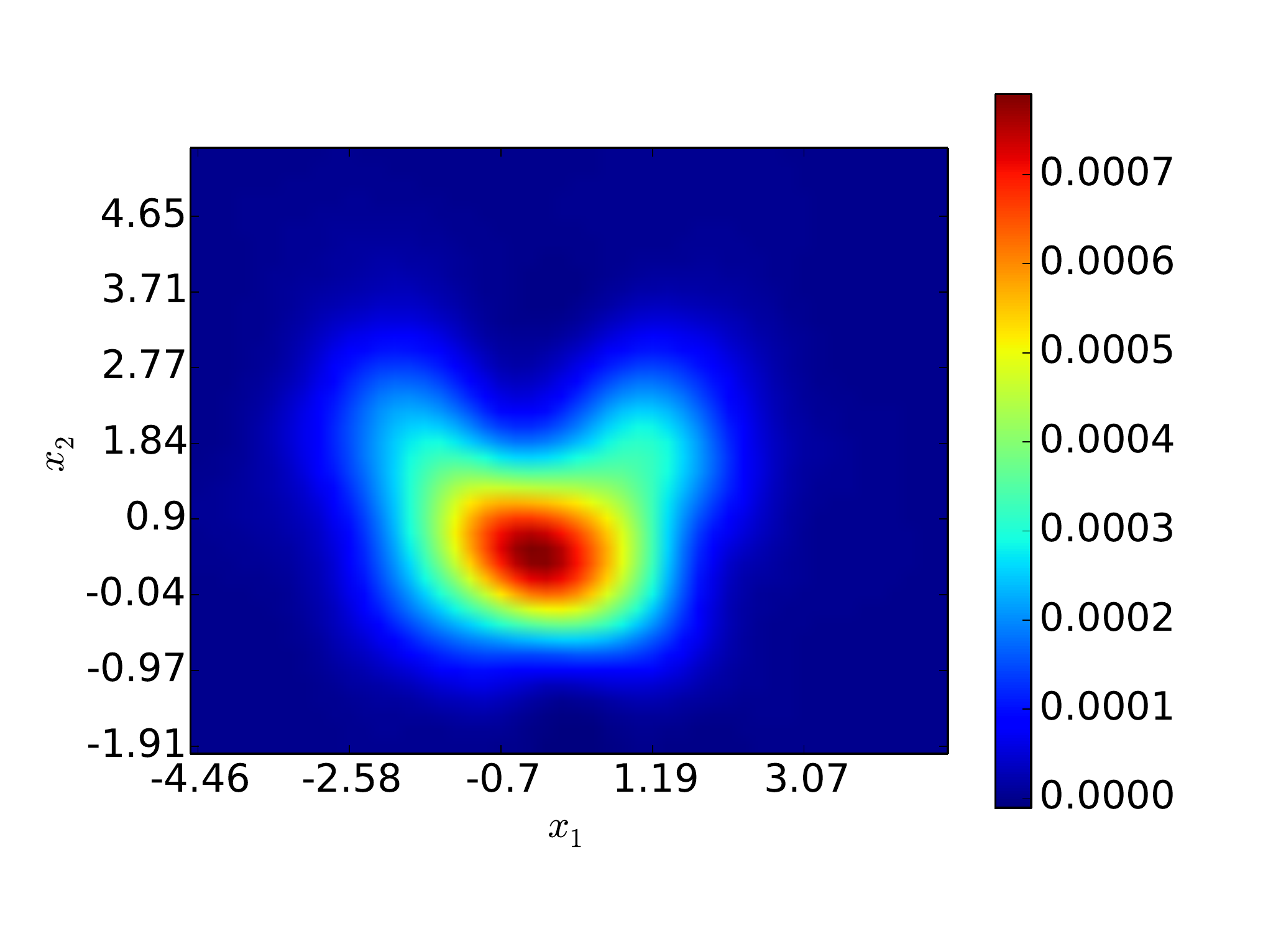}
            \caption{Zeroth (top), first (middle) and second (bottom) empirical (left) and estimated joint moments (right) with penalty weights $0.01$ for the zeroth moment, $0.05$ for the first moment and $0.1$ for the second joint moment.}
            \label{fig05}
        \end{figure*}

    \subsection{Joint moments}        

        As an example, this section demonstrates the application of the algorithm to $m^{(0)}$,~$m_1^{(1)}$, and $m_{1,1}^{(2)}$. 

To approximate the components of the joint moments, the procedure to select an appropriate penalty weight for the optimization layed out in Sec.~\ref{Moments} was carried out for $k \in \{1, 2, 3, 4, 5, 10, 20, 30, 40, 50\}$ and $\alpha^{(0)}, \alpha_i^{(1)}, \alpha_{i,j}^{(2)} \in \{0.001, 0.005, 0.01, 0.05, 0.1, 0.5, 1, 1.5, 2, 2.5, 3, 3.5, 4, 4.5,$ $5\}$. Figure \ref{fig04} shows that a penalty weight of $\alpha^{(0)} = 0.01$ and $\alpha_1^{(1)} = 0.05$ should be selected. For the second moment, the curves show a similar behaviour as for the first moment and a weight of $\alpha_{1,1}^{(2)} = 0.1$ is the appropriate choice (data not shown). To verify that the suggested measures lead to good values of the penalty weights, for each weight the squared sum of the differences between the estimated and the empirical joint moment $H^{(0)}$ and $H_i^{(1)}$ is computed. Results show, that the penalty weights selected with the method introduced above are a good choice, see insets fig.~\ref{fig04}. 

        The results of the optimization with the penalty weights determined as described above are shown in fig.~\ref{fig05}. As one sees, the estimates are smooth and fit well the original data, even in some of its small details.
        \begin{figure}[htb]
            \includegraphics[width = 0.5\textwidth]{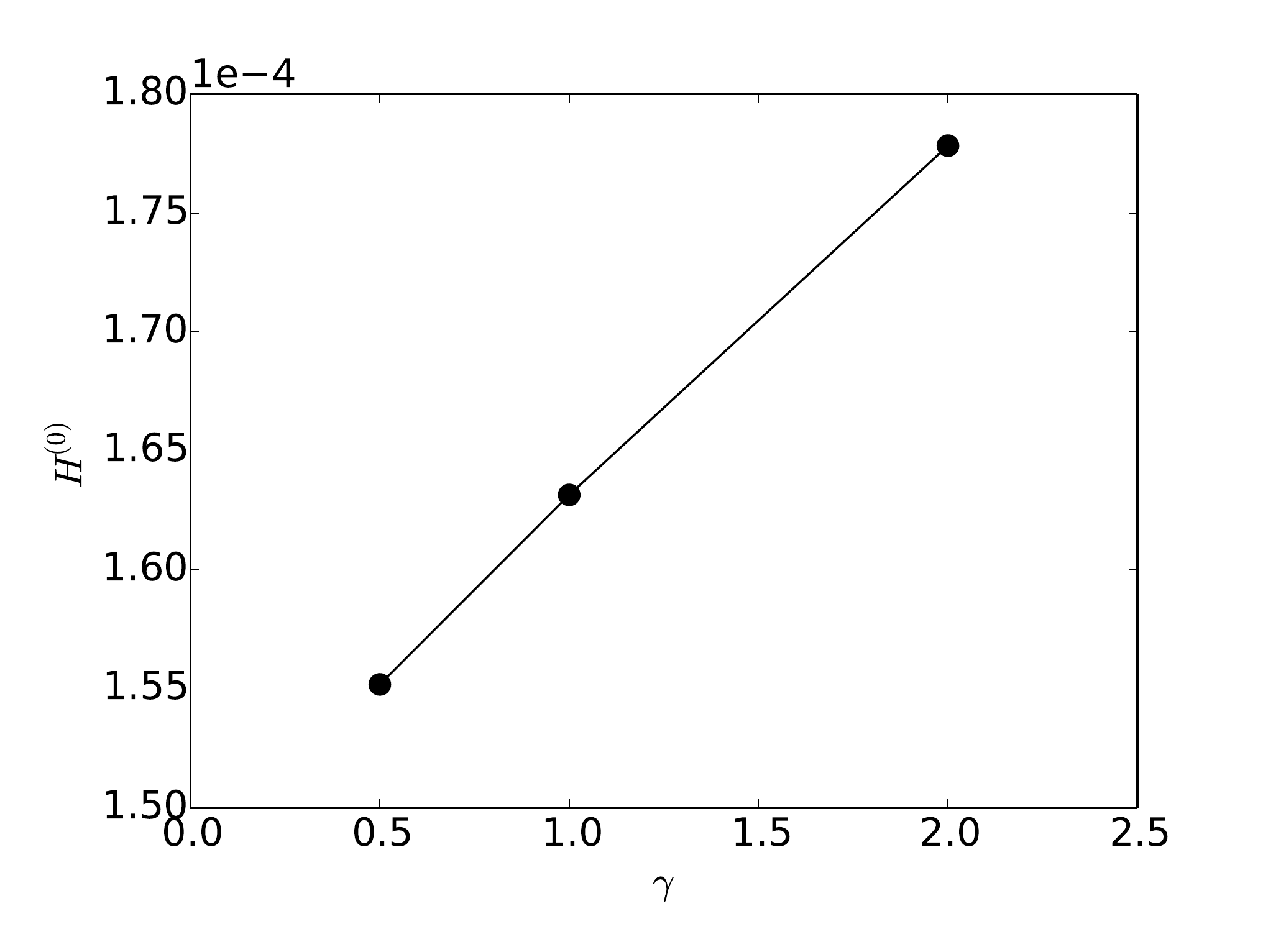}
            \includegraphics[width = 0.5\textwidth]{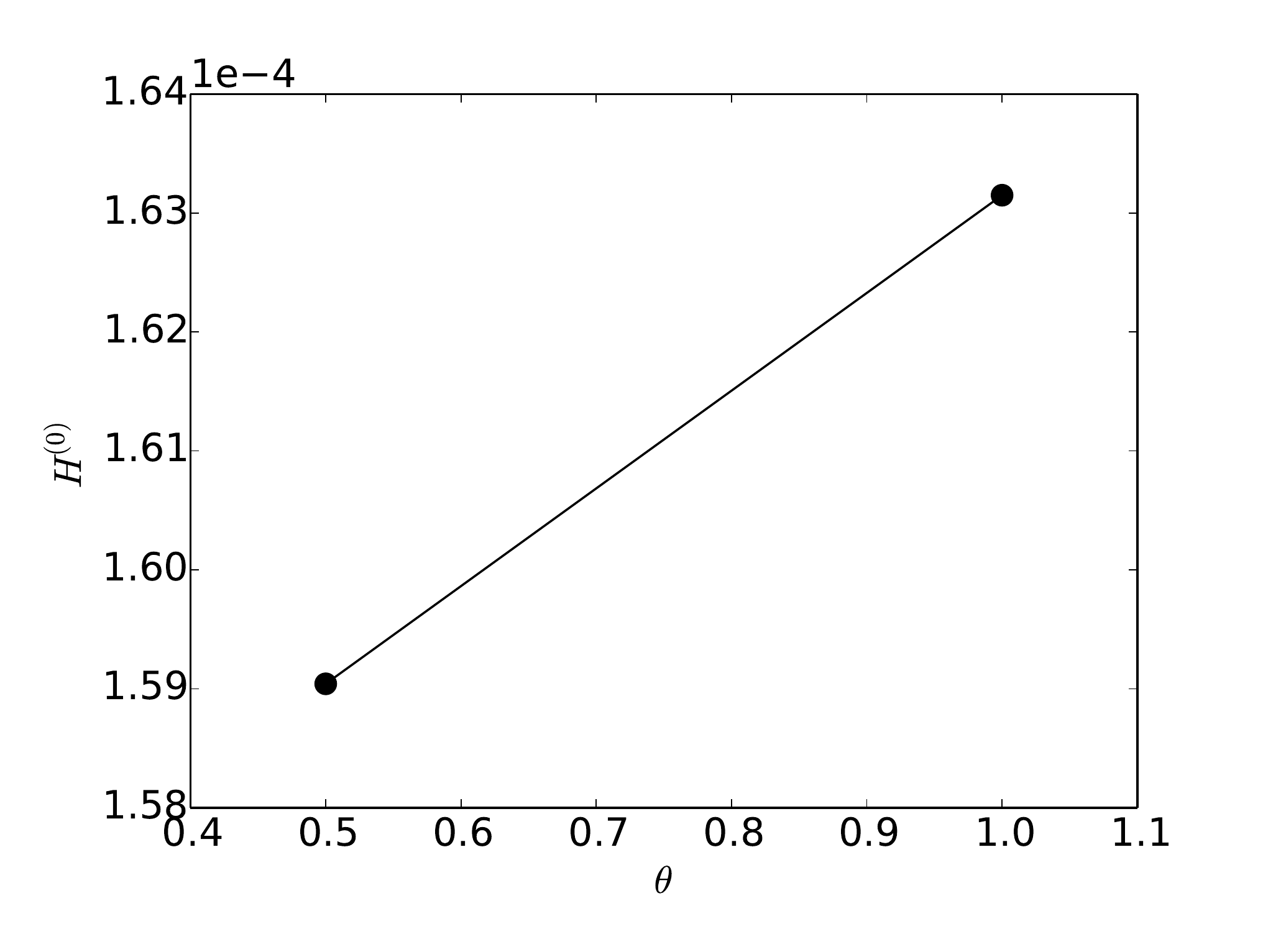}
            \caption{First component of the approximated (top) and analytical (bottom) drift coefficient.}
            \label{fig06}
        \end{figure}
        \begin{figure}[htb]
            \includegraphics[width = 0.5\textwidth]{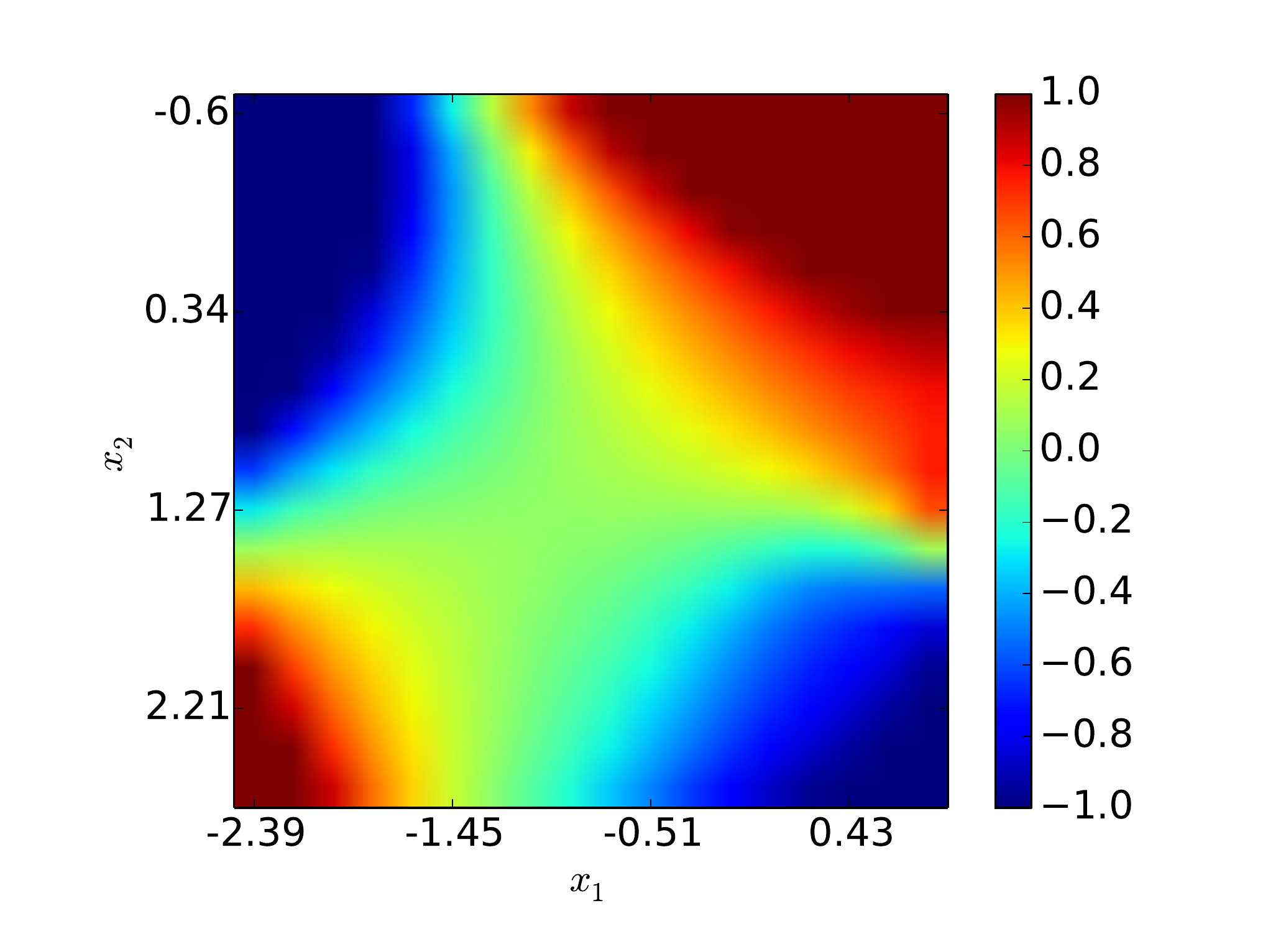}
            \includegraphics[width = 0.5\textwidth]{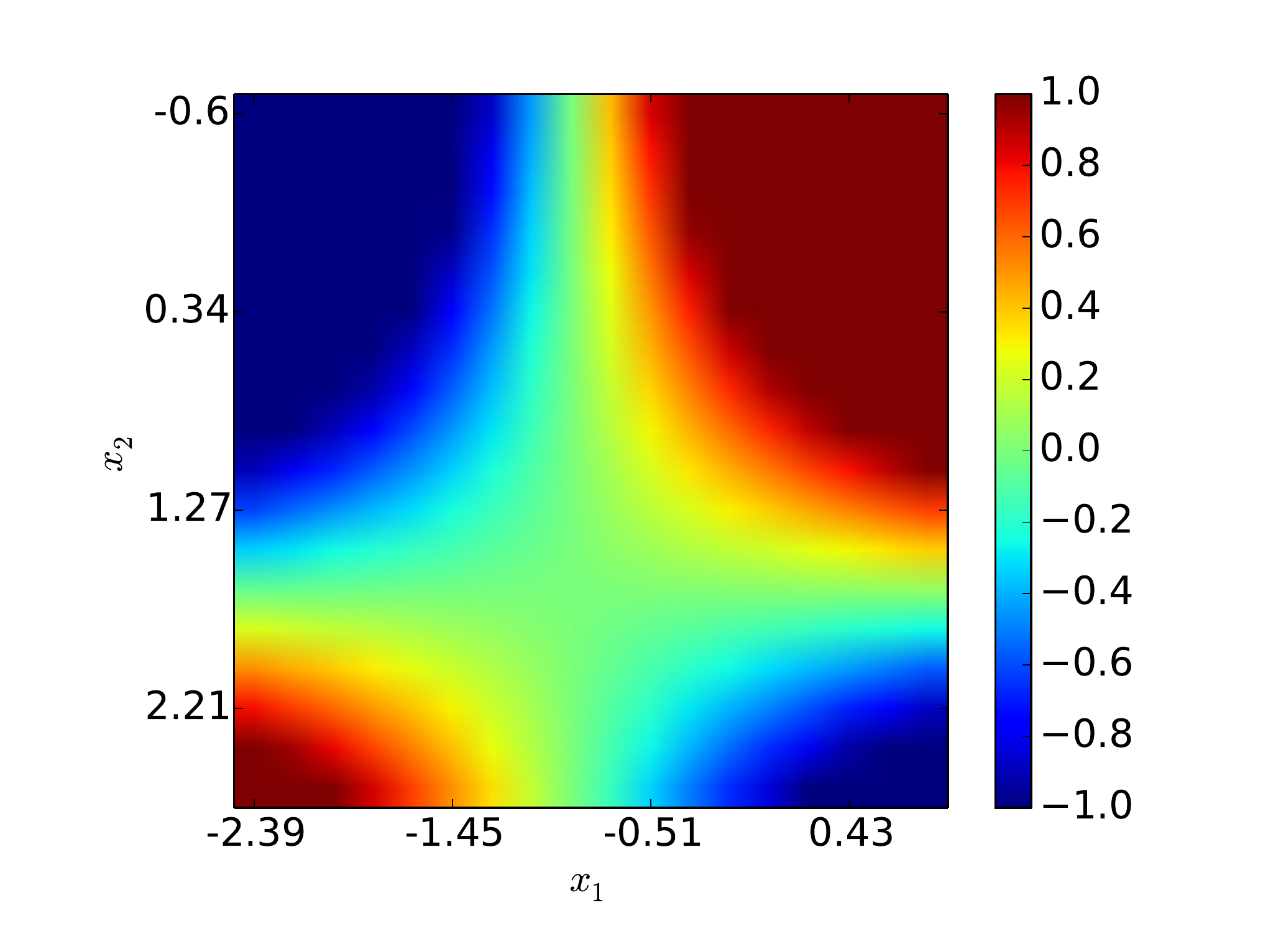}
            \caption{Distance $H^{(0)}$ of the approximated and the empirical zeroth joint moment for different measurement noise time-series generated with the parameter matrices $\mathbf{A}$ and $\mathbf{B}$ scaled by $\gamma$ and $\theta$, respectively. To vary the standard-deviation, values of $\theta = 0.5,1,2$ were chosen while $\gamma$ was constant $1$ (top). To vary the time-scale, both factors were set to $0.5$ and $1$ (bottom).}
            \label{fig07}
        \end{figure}
        
        From the approximated joint moments, using Eqs.~(\ref{eq6}), the drift and diffusion coefficients were estimated. 
        Figure \ref{fig06} presents the estimated (top) and analytical (bottom) first component of the drift coefficient for $x_1$-values between $-5$ and $4$ and $x_2$-values between $-2$ and $5$. Furthermore, the first component of the diffusion coefficient whose analytical value is constantly $0.5$ in all discretization bins was computed. Good results were obtained, however, as for the drift coefficient, the values on the margins were worse due to unsufficient data in the corresponding bins (data not shown). Thus, the presented method provides satisfactory results for the drift- and diffusion-coefficients of time-series data spoiled with strong measurement noise.        
        
    \subsection{Limits of the method}

        The previous section demonstrates that the presented method is able to extract good numerical estimations of the joint moments from a given stochastic process subject to measurement noise, see fig.~\ref{fig05}. To test the limits of this approach, several measurement noise time-series with different statistical properties were created. The two properties tested here are the variation of a) the standard deviation and, b) the time scale of the measurement noise. The proposed method was applied to the generated data and the results are presented and discussed in this section.
        
        The covariance matrix $\hbox{Cov}(\mathbf{Y})$ of a stationary, two-dimensional Ornstein-Uhlenbeck process $\mathbf{Y}(t)$, described by Eq.~(\ref{eq3prime}), is given by
            \begin{eqnarray}
                \hbox{Cov}(\mathbf{Y}) &=& 
                         \frac{\vert \mathbf{A} \vert
                           \sqrt{\mathbf{B}} \sqrt{\mathbf{B}}^T 
                          + \mathbf{\tilde{A}}
                              \sqrt{\mathbf{B}} \sqrt{\mathbf{B}}^T 
                             \mathbf{\tilde{A}}^T}
                        {2 Tr(\mathbf{A}) \vert \mathbf{A} \vert},
            \end{eqnarray}
        where $\mathbf{\tilde{A}}=(\mathbf{A} -
        Tr(\mathbf{A})\mathbf{Id})$ and $Tr(\mathbf{A})$ denotes the trace of $\mathbf{A}$ \cite{Gardiner2009}. Therefore, to generate time-series that vary in standard-deviation, not in time-scale, it suffices to scale the matrix $\mathbf{B}$ and keep the matrix $\mathbf{A}$ unchanged. To vary the time-scale, but not the standard-deviation, it is necessary to scale both matrices $\mathbf{A}$ and $\mathbf{B}$ with the same factor.
           
        For this analysis, four additional time-series $\mathbf{Y}_{(\gamma, \theta)}(t)$ were created, where the measurement noise parameters ${A}$ and $\mathbf{B}$ were scaled with factors $\gamma$ and $\theta$, respectively. The tested scaling factors were $(1, 0.5)$ to investigate the effects of a lower and $(1, 2)$ for a higher standard-deviation than the example presented in section~\ref{ExampleData}. To investigate the limit of the presented method in terms of time-scale, new measurement noise data was generated using scaling factors of $(0.5, 0.5)$ and $(2, 2)$ to yield a slower and a faster process than the section~\ref{ExampleData} example data, respectively. The generated measurement noise time-series were then added to the stochastic process and the proposed method applied the resulting noisy data. 

        Figure \ref{fig07} presents the summed square differences between the approximated and empirical zeroth joint moment $H^{(0)}$ in terms of variance [fig~\ref{fig07}(top)] and time-scale [fig.~\ref{fig07}(bottom)]. As expected, the error of approximation increases with increasing standard-deviation of the noise process. However, even for a ratio $\frac{\sigma_{noise}}{\sigma_{process}}$ of $0.75$ in the first component and $0.96$ in the second, i.e.~fluctuations in the measurement noise that are almost as big as the ones in the underlying stochastic process, satisfactory results can be obtained.

        In terms of time-scale, a slower measurement noise process yields better results [fig.~\ref{fig07}(bottom)]. For the noisy process on a faster time-scale than the example of section~\ref{ExampleData}, no results are presented, since the approximation of the time-correlation matrix $\mathbf{M}$ through equation system~(\ref{eq8}) was unsucessfull, likely, in this case the fast process could not be captured with the sampling time $\Delta t$. A more detailed discussion of the relation between the time-scale of the stochastic processes and the sucess of the approach is provided by Lehle~\cite{Lehle2013}.

\section{Conclusions}
\label{Conclusions}

    In conclusion, this paper investigated the possibility to infer the underlying stochastic process and the properties of measurement noise from a $N$-dimensional measured time-series. The presented approach is based on only three assumptions: a) that the process can be modeled as an Ornstein-Uhlenbeck process, b) that it operates on a faster time-scale than the stochastic process it superimposes and, c) that the two processes are independent.

    Moreover, it is parameter-free and thus can be applied to any Markovian multiplicative Gaussian white noise process. Application to synthetic data shows, that the presented method works for a wide range of amplitude ratios of the stochastic process and the measurement noise. In addition, the reconstruction succeeds with high accuracy for different time-scales of the measurement noise process.

    The algorithm requires solely a standard PC without special software environment and solves the inversion problem within the order of minutes. In the future, it will be applied to various problems that result from the mixing of stochastic processes, such as aerodynamic lift and drag measurements \cite{Luhur2014}.

\section*{Acknowledgements}

    The authors thank Funda\c{c}\~ao para a Ci\^encia e a Tecnologia for financial support under PEst-OE/FIS/UI0618/2011, PEst-OE/MAT/UI0152/2011, FCOMP-01-0124-FEDER-016080, SFRH/BPD/65427/2009 (FR) and SFRH/BD/86934/2012 (TS). PGL thanks the German Environment Ministry for financial support (0325577B). TS thanks the IPID4all programme supported by the German Academic Exchange Service (DAAD) with funds from the Federal Ministry of Education and Research (BMBF). This work is part of a bilateral cooperation DRI/DAAD/1208/2013 supported by FCT and DAAD.

\begin{appendix}

\section{Computation of B from V}  
    \label{appendixA}

        The measurement noise parameter $\mathbf{B}$ can be obtained from Eq.~(\ref{eq9b}) through partial integration. 
        \begin{widetext}
            \begin{eqnarray}
                \mathbf{V} &=& \int_0^{\infty}{e^{-\mathbf{A}s}\mathbf{B}e^{-\mathbf{A}^Ts}}ds\\
                           &=& -\left[ e^{-\mathbf{A}s} \mathbf{B} e^{-\mathbf{A}^Ts} (\mathbf{A}^T)^{-1}\right]_0^{\infty} - \int_0^{\infty}{ \mathbf{A} e^{-\mathbf{A}s} \mathbf{B} e^{-\mathbf{A}^Ts}(\mathbf{A}^T)^{-1}}ds\\
                           &=& -\lim \limits_{s \to \infty} \left[e^{-\mathbf{A}s} \mathbf{B} e^{-\mathbf{A}^Ts} (\mathbf{A}^T)^{-1}\right] + \mathbf{B}(\mathbf{A}^T)^{-1} - \left[ \mathbf{A} \int_0^{\infty}{e^{-\mathbf{A}s} \mathbf{B} e^{-\mathbf{A}^Ts}}ds (\mathbf{A}^T)^{-1}\right]\\
                           &=& \mathbf{B}(\mathbf{A}^T)^{-1} - \mathbf{AV}(\mathbf{A}^T)^{-1}
            \end{eqnarray}
        \end{widetext}

\end{appendix}

\bibliography{MeasNoiseBib.bib}

\end{document}